\documentclass[12pt]{article}
\usepackage{amsfonts}
\usepackage{bbm}
\usepackage{bm}
\usepackage{amsmath}
\usepackage{amscd}

\usepackage{amssymb,lscape}
\usepackage[matrix,arrow,curve]{xy}

\newcommand{\bmat}{\left(\begin{array}}
\newcommand{\emat}{\end{array}\right)}

\def\gtrsim{\mathrel{\raise.3ex\hbox{$>$\kern-.75em\lower1ex\hbox{$\sim$}}}}

\def\-{\hphantom{-}}

\def\s2{\frac{1}{\sqrt2}}

\def\beq{\begin{equation}}
\def\eeq{\end{equation}}
\def\beqa{\begin{eqnarray}}
\def\eeqa{\end{eqnarray}}

\def\bom{{\bm\omega}}
\def\bQ{{\bm Q}}
\def\bP{{\bm P}}

\def\im{{\rm Im \,}}
\def\re{{\rm Re \,}}

\def\T{{\rm T}}
\def\Z{{\mathbb Z}}

\def\cj{{\cal J}}

\def\mg{m_{3/2}}
\def\mg2{m^2_{3/2}}

\def\Dsl{\,\raise.15ex\hbox{/}\mkern-13.5mu D} 
\def\rep#1{\mbox{{\bf #1}}}

\def\bom{{\bm\omega}}

\newcommand{\bbR}{\mathbb{R}}

\DeclareMathOperator{\SL}{\mathit{SL}}
\DeclareMathOperator{\GL}{\mathit{GL}}

\DeclareMathOperator{\E7}{\mathit{E}_{7}}

\newcommand{\Tsub}{O(6,6) \times \SL(2,\bbR)}
\newcommand{\Tsubh}{O(6,6) \times \SL(2,\bbR)|_H }
\newcommand{\Tsubb}{O(6,6) \times \SL(2,\bbR)|_{B} }
\newcommand{\SLR}{\SL(2,\bbR)}

\newcommand{\mukai}[2]{\big<{#1},{#2}\big>}

\newcommand{\id}{\mathbb{I}}
\newcommand{\eq}[1]{(\ref{#1})}
\newcommand{\Ggeom}{G_{\mathrm{geom}}}
\newcommand{\nn}{\nonumber}

\newcommand{\bj}{{\bar\jmath}}

\newcommand{\tj}{{\tilde\jmath}}

\newcommand{\bbG}{\mathbb{G}}

\begin{document}
\pagestyle{plain}

\makeatletter
\@addtoreset{equation}{section}
\makeatother
\renewcommand{\theequation}{\thesection.\arabic{equation}}
\pagestyle{empty}
\vspace{5mm}
\rightline{CERN-PH-TH-2010-172}
\vspace{-2mm}
\rightline{IPhT-T10/093}
\begin{center}
\vspace{12mm}
\LARGE{\bf U-dual fluxes and Generalized Geometry\\[11mm]}
\large{\bf  G. Aldazabal${}^{a,b}$, E. Andr\'es${}^{b}$, P. G. C\'amara${}^c$ and M. Gra\~na${}^d$
 \\[4mm]}
\small{
${}^a${\em Centro At\'omico Bariloche,} ${}^b${\em Instituto Balseiro
(CNEA-UNC) and CONICET.} \\[-0.3em]
{\em 8400 S.C. de Bariloche, Argentina.}\\
[0.3cm]
${}^c${\em CERN, PH-TH Division, 
CH-1211 Gen\`eve 23, Switzerland.}\\
[0.3cm]
${}^d${\em Institut de Physique Th\'eorique,
CEA/ Saclay \\
91191 Gif-sur-Yvette Cedex, France}  \\
[0.4cm]}
\small{\bf Abstract} \\[0.5cm]
\end{center}

{\small We perform a systematic analysis of generic string flux compactifications, making use of Exceptional Generalized Geometry (EGG) as an organizing principle. In particular, we establish the precise map between fluxes,
gaugings of maximal 4d supergravity and EGG, identifying the complete set of gaugings that
admit an uplift to 10d heterotic or type IIB supegravity backgrounds. Our results reveal a rich structure, involving new deformations of 10d supergravity backgrounds, such as the RR counterparts of the $\beta$-deformation. These new deformations are expected to provide the natural extension of the $\beta$-deformation to full-fledged F-theory backgrounds. Our analysis also provides some clues on the 10d origin of some of the particularly less understood gaugings of 4d supergravity. Finally, we derive the explicit expression for the effective superpotential in arbitrary $\mathcal{N}=1$ heterotic or type IIB orientifold compactifications, for all the allowed fluxes.}

\newpage
\setcounter{page}{1}
\pagestyle{plain}
\renewcommand{\thefootnote}{\arabic{footnote}}
\setcounter{footnote}{0}

\tableofcontents

\section{Introduction}
\label{seci}
T-duality is a distinctive symmetry of String Theory, in the sense that stringy aspects  come into play. In its simplest form it identifies a theory  compactified  on a circle of a given radius with  a theory compactified on a circle of inverse radius  and exchanges compact momenta and winding modes, which simply do not exist  in a field theory of particles.
 When compactification on a more general  background with $d$ isometries is considered,  T-duality action is enhanced to an $O(d,d,\Z)$ group that, among other features,   mixes the metric modes with the antisymmetric NSNS $B_2$ field components.
In a 2d sigma model approach, Buscher rules \cite{buscher} indicate
the precise  way in which target space fields  transform under T-duality, highlighting the fact that  different backgrounds lead to the same CFT.

When fluxes are turned on \cite{grana}, the situation becomes richer and more intricate.
A clear illustration is provided by a torus compactification in the presence of a flux of the NSNS 3-form  $H_3=dB_2$. It has been suggested that, taking  T-duality symmetry  as a fundamental symmetry of String Theory
would lead to include new ``fluxes'',  following the rules
\begin{equation} \label{Tdualchain}
H_{mnp}\ {\stackrel{\T_m}
{\longleftrightarrow}} \ \omega^m_{np}\ {\stackrel{\T_n}
{\longleftrightarrow}} \ Q^{mn}_p\ {\stackrel{\T_p}
{\longleftrightarrow}}\ R^{mnp}\ .
\end{equation}
where $T_m$ denotes a T-duality transformation performed along an internal direction $m$ \cite{stw,hull,acfi}.
Such indication manifests by comparing 4d effective superpotentials derived from orientifold compactifications of the type IIA and IIB 10d supergravity actions. Components in (\ref{Tdualchain}) are related to the various coefficients of effective superpotential couplings in the dimensionally reduced theory. Since the corresponding type IIA and IIB string theories are supposed to be connected by mirror symmetry, these new fluxes must be included in order for the superpotentials to match.

An obvious question arising  at this stage concerns
the higher dimensional origin of these new fluxes. The first transformation in (\ref{Tdualchain}) is well understood from Buscher rules. If a T-duality transformation is performed along the isometry direction $m$ ($B_2$  being independent of this direction), we end up with a background with no $H_3$ flux, corresponding  to a compactification of 10d supergravity on a different manifold, named  twisted torus. This is characterized by the first Chern class  of the spin bundle, given by  $\omega^m_{np}$.

By performing a new duality, let us say along $n$, a new 10d supergravity background is obtained,\footnote{Strictly
speaking, the usual derivation of Buscher rules is actually not
allowed in these cases since the isometry is not globally well
defined. However, following these rules even in these ``obstructed'' cases seems to be meaningful.
For instance, to some of these solutions an interpretation in terms of asymmetric orbifolds can be given \cite{Hull:2006qs} (see also \cite{marcheschul}).} which can be
locally characterized by the tensor $Q^{mn}_{p}=\partial_p\beta^{mn}$. The new ($\beta$-deformed) background, however, does not define a global manifold in the usual sense of Riemannian geometry. In order to link such solutions in different intersecting coordinate patches, local solutions can be connected by the usual geometric transition functions involving diffeomorphisms and gauge transformations. However, after a global transformation solutions are connected only if T-duality transformations are also  allowed for as part of the transition functions. For that reason $Q^{mn}_p$ is also called a non-geometric flux.

The last T-duality transformation in eq.(\ref{Tdualchain}) is more obscure. In fact, it has been conjectured \cite{stw} from the 4d effective action viewpoint that if a T-duality transformation along a direction $p$ is performed in a background with $Q^{mn}_p$ fluxes, we would arrive to a ``truly non-geometric'' flux $R^{mnp}$ for which not even a local 10d supergravity description is available. In this sense, the resulting 4d effective supergravity should  not be thought of  as  a dimensional reduction of a 10d effective supergravity background,  but as a 4d theory that incorporates  information of the full string theory.

From the above chain of T-dualities a ``stringy'' generalization of the concept of Riemannian  geometry seems to emerge, on which symmetries of the $B_2$ field are taken now on equal footing than diffeomorphisms. Thus, in a series of papers starting from \cite{GMPT,JW}, the concept of  Generalized Geometry \cite{hg} (see also \cite{koerber} for an introduction to the topic) was proposed as a natural framework to describe string compactifications with fluxes.
In Generalized Geometry the full T-duality group $O(6,6)$ ($O(d,d)$ for compactification on $d$ dimensions) is the structure group of a generalized bundle built up from the tangent and cotangent bundles of the six ($d$) dimensional manifold.
Namely, vectors in this generalized bundle are built up from vectors and one-forms of the original one.
The generalized metric combines  the original metric and the $B_2$ field. Patching in overlapping regions requires, besides diffeomorphisms, an $O(6,6)$ action involving, in particular, the $B_2$ field.\footnote{It is worth noticing that, in this description, fields do  depend on six dimensional coordinates  of the original compactification space. A more ambitious program points towards  a geometric description where fields depend on coordinates on a ``doubled torus'' \cite{dt1,dt2} in twelve  dimensional space.}

Hence, Generalized Geometry is particularly powerful in dealing with general 4d $\mathcal{N}=2$ and $\mathcal{N}=1$ compactifications. The existence of a single nowhere vanishing spinor requires the local $O(6,6)$ structure to be reduced to a \emph{global} $SU(3)$ structure, characterized in terms of a globally $SU(3)$ invariant K\"ahler (1,1)-form $J$ and a holomorphic 3-form $\Omega$. The 10d supergravity equations of motion can be then recast in terms of F-term and D-term densities depending on $J$ and $\Omega$, which after Kaluza-Klein reduction and integration over the compact space give rise to the F-terms and D-terms of the  effective 4d gauged supergravity theory \cite{GLW,BG}.

The above picture, however, is far from being complete. By invoking other duality transformations of String Theory, like IIB S-duality, M-theory or heterotic/type I S-duality extra fluxes are suggested  \cite{acfi}. Again, such new fluxes can be inferred by matching couplings in 4d effective superpotentials. Similarly, when dealing with  general gauged supergravity theories (see \cite{samt} for a review) gauge symmetries and structure constants can be accounted for in a string theory framework only if these new fluxes, associated to obstructed dualities, are incorporated. In this regard, an extension of the framework of Generalized Geometry is called for.

The natural generalization appears to be Exceptional (or Extended)  Generalized Geometry (EGG) \cite{HullEGG,EGG}. In EGG the structure group of the generalized bundle is now the full U-duality group, $E_7$, so that symmetries of the RR fields are also naturally incorporated. The effect of the orientifold projection on the untwisted modes can be then understood from the breaking of the local $E_7$ structure to some $O(6,6)\times SL(2)$ subgroup.

The main aim of this paper is to use the tools of EGG  as an organizing principle for generic string flux compactifications. In the first part of the paper we focus on aspects which are directly related to the local $E_7$ structure of the compactification. In particular, we perform a complete mapping between fluxes, gaugings of 4d gauged supergravity and EGG. This allows for a systematic determination of all the gaugings of 4d gauged supergravity that admit an uplift to backgrounds of 10d supergravity.\footnote{Related partial results can be found in \cite{dauria,ferrara,ferrara2,bhk2,acr,Dall'Agata:2009gv}.} These are summarized in eqs.(\ref{h1})-(\ref{ps}) of Section \ref{sec2b}. Apart from already known deformations, such as backgrounds for the RR and NSNS field-strengths of 10d supergravity, metric fluxes or $\beta$-deformed backgrounds, our analysis also reveals new deformations related to the RR counterparts of the $\beta$-deformation. These provide the natural extension of the latter to full-fledged F-theory backgrounds (see also \cite{acr}). Moreover, our analysis also sheds light on the 10d origin of some of the particularly less understood gaugings of 4d $\mathcal{N}=4$ supergravity, such as the ones transforming in  the vector representation of $O(6,6)$ \cite{sw} (see also \cite{vzminimal,marios,Dall'Agata:2009gv} for related work).

In the second part of the paper then we consider aspects related to the global $SU(3)$ structure of $\mathcal{N}=1$ compactifications. More precisely, making use of the tools of EGG, we write explicitly the effective superpotential for all the U-dual fluxes allowed in general $\mathcal{N}=1$ heterotic or type IIB orientifold compactifications.

A more detailed outline of the paper is as follows. In Section \ref{sec0} we introduce the basic concepts of Generalized Geometry. In particular we discuss the action of different generators of $O(6,6)$ structure group in connection  with diffeomorphisms, $B_2$ gauge transformations and the extra $\beta$ transformation, associated to non globally defined geometries. The interpretation of NSNS field strength fluxes associated to such transformations and how they combine to fulfill an $O(d,d)$ representation is also presented.
In Section \ref{sec2} we address  the inclusion of RR fields into geometry, defining EGG.  The structure group is thus enlarged to the full $E_7$ U-duality group and the generalized tangent space becomes 56 dimensional.
U-dual field strengths  are discussed in Section \ref{sec2b}. In Section \ref{sectori} we consider the simplest compactifications, that is toroidal compactifications, in the presence of general dual fluxes. We discuss the algebra and global constraints that U-dual fluxes must satisfy, and derive the flux induced effective superpotential in these compactifications. The relation with 4d $\mathcal{N}=8$ and $\mathcal{N}=4$ gauged supergravity is also discussed in that Section.
Section \ref{sec4} is devoted to the formulation of  $\mathcal{N}=1$ backgrounds in EGG and to the construction of $\mathcal{N}=1$ untwisted sector superpotentials for general $SU(3)$ structure compactifications. Section  \ref{sec5} provides some final comments. Some notation and useful results related to the $E_7$ algebra are summarized in the Appendices.

\section{Generalized Geometry}
\label{sec0}

\subsection{$O(d,d)$ structure}

In Generalized Geometry \cite{hg}, the T-duality group $O(d,d)$ is  the structure group of a generalized bundle constructed by combining the tangent $TM$ and cotangent $T^* M$ bundles of a $d$-dimensional manifold $M$.
A generalized metric on this generalized bundle encodes information about the metric and the $B_2$ field of the manifold. Matching of overlapping patches is achieved by allowing  transformations involving the $B_2$ field besides the usual diffeomorphisms. In that way the $B_2$ field is incorporated to the geometry of the generalized bundle.

At the intersection of two patches $U_\alpha$ and $U_\beta$, the generalized vectors $X=x+\xi$ (where $x$ is a vector and $\xi$ a one-form), are identified as follows
\beq \label{patch1}
x_\alpha+\xi_\alpha = a_{\alpha \beta} x_\beta + (a^{-T}_{\alpha \beta} \xi_\beta -\iota_{x'_\beta} b_{\alpha\beta}) \ ,
\eeq
where  $a \in GL(d,{\mathbb R})$ (and $a^{-T}\equiv(a^{-1})^T$) is a conventional diffeomorphism, $b_{\alpha\beta}$ is a two-form and $\iota_{x'_\beta}$ means a contraction along $x'_\beta \equiv a_{\alpha \beta} x_\beta$. This patching corresponds to the following transformation $h \in O(d,d)$
\begin{eqnarray}
\label{eq:patch}
   X_{\alpha}
      = \begin{pmatrix} x_{\alpha} \\ \xi_{\alpha} \end{pmatrix}
      &=&  \begin{pmatrix}
            \id & 0 \\  b_{\alpha\beta} & \id
         \end{pmatrix}
         \begin{pmatrix}
            a_{\alpha\beta} & 0 \\ 0 & a^{-T}_{\alpha\beta}
         \end{pmatrix}
         \begin{pmatrix} x_{\beta} \\ \xi_{\beta} \end{pmatrix} \\
         &\equiv& \quad \quad e^{\cal B} \quad \quad    \begin{pmatrix}
            a_{\alpha\beta} & 0 \\ 0 & a^{-T}_{\alpha\beta}
         \end{pmatrix}\begin{pmatrix} x_{\beta} \\ \xi_{\beta} \end{pmatrix}
      = h_{\alpha\beta} X_{\beta} \,  \nn
\end{eqnarray}
where ${\cal B}=  \begin{pmatrix}
            0 & 0 \\  b_2 & 0         \end{pmatrix}
$ is the $O(d,d)$ generator corresponding to an Abelian subgroup, $G_{\cal B}$.

The presence of the 2-form $b_2$ in the patching implies that the generalized tangent bundle is not just the sum $TM\oplus T^*M$ but is an extension of the tangent bundle by the cotangent one
\begin{equation}
\label{eq:Edef}
   0 \longrightarrow T^*M \longrightarrow E
      \stackrel{\pi}{\longrightarrow} TM \longrightarrow 0 \ ,
\end{equation}
where $b_{\alpha\beta}$ describes how $T^*M$ is fibered over $TM$.  $b_2$ is required to be locally exact, i.e. $b_{\alpha\beta}=d\Lambda_{\alpha\beta}$ and (in an analogous way as one patches a $U(1)$ bundle) the local 1-forms $\Lambda_{\alpha\beta}$ should satisfy the cocycle condition
\begin{equation}
\label{cocycle}
   \Lambda_{\alpha\beta} + \Lambda_{\beta\gamma}
         + \Lambda_{\gamma\alpha}
      = g_{\alpha\beta\gamma} d g_{\alpha\beta\gamma}
\end{equation}
on triple intersections $U_\alpha \cap U_\beta \cap U_\gamma$, where $g_{\alpha\beta\gamma}=e^{i \theta} \in U(1)$. The gauge parameters $\Lambda_{\alpha\beta}$ allow to define a local 2-form gauge field $B_{\alpha}$ whose patching is
\beq \label{Bpatches}
B_{\alpha}=B_{\beta}+ d \Lambda_{\alpha\beta} \ ,
\eeq
and whose field strength $H_3=dB_2$ determines the quantized global curvature of the gerbe.

Diffeomorphisms and $b$-transforms in \eq{eq:patch} form a subgroup of $O(d,d)$ that is a semi-direct
product $\Ggeom=G_{\cal B} \rtimes GL(d)$. Specializing to the case $d=6$, this subgroup has 51 generators (36 generate $GL(6)$ and 15 make up $G_{\cal B}$).
As we will see in Section \ref{sec2}, they form a parabolic subgroup of $O(6,6)$. The remaining 15 generators of $O(6,6)$ define another Abelian subgroup
\begin{equation}
\label{eq:betatransform}
   e^{\beta_2} =
      \begin{pmatrix} \id & \beta_2 \\ 0 & \id \end{pmatrix} \ ,
\end{equation}
characterized by a bi-vector $\beta_2$. The ``$\beta$-transform" action on a generalized vector is  $   X = x+\xi \mapsto (x + \beta\llcorner\xi)+  \xi $ (in components, this is
$x^a + \xi_a \mapsto (x^a + \beta^{ab} \xi_b) + \xi_a$). One could a priori allow for patchings involving these transformations
as well. This is perfectly fine as long as $\beta_2$ is a globally defined bi-vector.
If that is not the case, there is no
gauge transformation like (\ref{Bpatches}) that allows to define it patchwise. Roughly speaking, this is because the derivative in (\ref{Bpatches}) has an index down, and even when combined with a vector it cannot define a bi-vector. This is related to the fact that the manifold  is still $d$-dimensional, even if we have defined an extended $2d$-dimensional tangent bundle on it.
Hence, whenever $\beta_2$ is not globally well-defined, there is no ``gerbe-like" construction such as that outlined above for $B_2$.

The 36-dimensional space of metric and $B_2$ field parameterize the coset $\frac{O(6,6)}{O(6) \times O(6)}$, or in other words $g$ and $B_2$ combine to define an $O(6) \times O(6)$ structure on $E$. This structure can also be defined by the splitting of the
generalized tangent bundle into two orthogonal 6-dimensional sub-bundles $E=C_+ \oplus C_-$ such that the natural $O(6,6)$-invariant metric
\beq
   \eta = \begin{pmatrix} 0 & \id \\ \id & 0 \end{pmatrix}
\eeq
decomposes into a positive-definite metric on $C_+$ and a negative-definite metric on $C_-$. The sub-group of $O(6,6)$ that
preserves each metric separately is $O(6) \times O(6)$. One can now define a positive definite generalized metric
\begin{equation}
   {\cal H} = \left.\eta\right|_{C_+} - \left.\eta\right|_{C_-} .
\end{equation}

A generic element of $C_+$ or $C_-$ cannot be purely a vector
or a one-form, since these are null with respect to $\eta$. We can therefore write $X_+ \in C_+$ as $x+M\, x$ for some matrix $M$. Taking
$M=B_2+g$, the patching condition  (\ref{patch1}) implies $g_{\alpha}=g_{\beta}$, while $B_2$ is patched according to \eqref{Bpatches}.
Orthogonality between $C_+$ and $C_-$ implies that an element $X_- \in C_-$ can be written as $X_-=x+(B_2-g) \, x$.
When we write  a generic element $X=x+\xi \in E$ as $X=X_+ + X_-$, with $X_\pm=x+(B_2\pm g) \, x$ we find that the generalized metric takes the form
\begin{equation}
\label{eq:genmetric}
   {\cal H} = \begin{pmatrix}
         g - B_2 g^{-1} B_2 & B_2 g^{-1} \\
         - g^{-1} B_2 & g^{-1}
      \end{pmatrix} .
\end{equation}
and parameterizes the space of scalar fields of the resulting 4d effective supergravity.
The generalized metric can be acted by $O(6,6)$ transformations. The effect of diffeomorphisms
is changing the metric and $B_2$ field accordingly (namely $(g,B_2) \to a^{T} (g,B_2) a$), while that of $b$-transformations is to shift $B_2\to B_2-b_2$. One can therefore think about a background with a given $B_2$ field $B$ as the $b_2$-transform with $b=-B$ of a background with no $B_2$ field. 

A $\beta$-transformation leads to a more complicated action so that it does not seem possible to
tell apart the new $g$
and the new $B_2$. However, it is always allowed
to perform an $O(6) \times O(6)$ transformation in the stabilizer of $g+B_2$ (i.e., a transformation that does not change the metric and $B_2$ field)
such that the $\beta$-transform rotates into a diffeomorphism and a $b$-transform.  In other words, given a metric and $B_2$ field at a  point on the manifold, the full $\frac{O(6,6)}{O(6) \times O(6)}$ orbit can be reached by acting only with the geometric subgroup of $O(6,6)$ that implies that $\beta$-transforms can be locally ``gauged away". We will come back to this point in the next Section.

The transformations $b_2$ and $\beta_2$  are related by T-duality. In fact, there is another basis for  $O(d,d)$ given by diffeomorphisms, $b$-transforms and T-dualities along any two directions on the tangent space. A T-duality along the first two basis vectors is realized by the following
$O(6,6)$ transformation
\begin{equation}
 \label{Tdual12}
   T_{12} = \begin{pmatrix}
         0 & 0 & \id_2 & 0 \\
         0 & \id_4 & 0 & 0\\
         \id_2 & 0 & 0& 0\\
         0 & 0 & 0 & \id_4
      \end{pmatrix} ,
\end{equation}
where $\id_{2,4}$ are the $2 \times 2$ and $4 \times 4$ identity matrices. Hence, T-duality transforms each fundamental $O(6,6)$ form $X_A=(\xi_f,\xi_b;v^f,v^b)$ into $(v^f,\xi_b;\xi_f,v^b;)$.

It is not hard to check that the action of T-duality along a fiber $f$ on a generic element (split among its base and fiber components) in the Lie algebra
$o(d,d)$ is
\beqa \label{TdualgenO66}
A^A{}_B=\begin{pmatrix}
        a^f{}_f & a^f{}_b &  \beta^{ff} &\beta^{fb} \\
       a^b{}_f & a^b{}_b &  \beta^{bf} &\beta^{bb} \\
    b_{ff} & b_{fb} &  \hat a_f{}^f & \hat a_f{}^b \\
    b_{bf} & b_{bb} &  \hat a_b{}^f & \hat a_b{}^b
      \end{pmatrix}
  \longrightarrow \, T_{f}^T \, A \, T_{f}&=& \begin{pmatrix}
        \hat a_f{}^f & b_{fb} &  b_{ff} &\hat a_f{}^{b} \\
        \beta^{bf}  & a^b{}_b & a^b{}_f&\beta^{bb} \\
  \beta^{ff}  & a^f{}_b &  a^f{}_f  &\beta^{fb} \\
  \hat a_b{}^f & b_{bb} & b_{bf}& \hat a_b{}^b
      \end{pmatrix}  \equiv \nn \\
      \tilde A &=& \begin{pmatrix}
        \tilde a^f{}_f & \tilde a^f{}_b &  \tilde \beta^{ff} &\tilde \beta^{fb} \\
      \tilde a^b{}_f & \tilde a^b{}_b &  \tilde\beta^{bf} &\tilde\beta^{bb} \\
   \tilde b_{ff} &\tilde b_{fb} &  \hat \tilde \hspace{-1mm} a_f{}^f & \hat\tilde \hspace{-1mm}a_f{}^b \\
   \tilde b_{bf} & \tilde b_{bb} &  \hat\tilde \hspace{-1mm} a_b{}^f & \hat\tilde \hspace{-1mm} a_b{}^b
      \end{pmatrix}  \ .
 \eeqa
Here the index $A=1,...,2d$ is split into $d$ indices up and $d$ indices down, and each of them is further split into fiber and base indices. We have also defined $\hat a\equiv -a^{T}$ and $b_2$ and $\beta_2$ are antisymmetric, i.e. $b_{fb}=-b_{bf}$, $\beta^{fb}=-\beta^{bf}$. We see that the effect of T-duality is, roughly speaking, to raise and lower $f$-indices and move
the building block to the corresponding new place according to the structure of the indices. Hence, transformations purely on
the base are not touched by T-duality. A fiber-base (base-fiber) diffeomorphism is exchanged with a fiber-base (base-fiber) $b_2$ ($\beta_2$)-field,
while $b_2$ and $\beta_2$ purely along the fiber are interchanged. A $\beta$-transform
with one or two legs along the fiber can therefore be generated by T-dualizing a diffeomorphism or a $b_2$ field with respectively one and two legs along the fiber.

\subsection{T-dual field strengths}
\label{sec1}

Analogous to the local definition of $H_3$, we can also introduce some field strengths for all the other $O(d,d)$ transformations,\footnote{Note that these are all local definitions. Global aspects will be mostly discussed in Sections \ref{sectori} and \ref{sec4}.}
\beq
H_{mnp}=\partial_{[m} b_{np]} \ , \quad \omega_{mn}^p=\partial_{[m} a_{n]}{}^p \ , \quad Q_m^{np}=\partial_m \beta^{np} \ .\label{tfield}
\eeq
There are 20 components of $H_3$, while $\omega$ and $Q$ have 90 components
each. This gives a total of 200 components. The smallest $O(6,6)$ representation containing 200 elements is the $\rep{220}$, which consists of
3-forms $F_{ABC}$ on $E$. The $O(6,6)$  index $A$ of the fundamental representation can be split into six indices $m$ down and six indices up (i.e. a form and a vectorial index). To fill out the $\rep{220}$ representation  an object with three indices up is missing. Such ``locally non-geometric" flux, needed to restore T-duality covariance of the 4d low energy action, has been termed $R$-flux \cite{stw}. It is the candidate T-dual of $H_{mnp}$ along $m, n$ and $p$ (though such T-duality \`a la Buscher \cite{buscher} is forbidden since in order to get a flux $H_{mnp}$ from $B_{np}$, the latter should depend on the coordinate $x^m$, which is therefore not an isometry).  From the point of view of representations of
$O(d,d)$, we can see that the $O(6,6)$ transformation corresponding to three T-dualities along $mnp$ acting on the element $H_{mnp}$ of the 220 representation gives indeed another element with three indices up.
This generates the chain in (\ref{Tdualchain}).
As for the local non-geometricity, it can be easily appreciated that
a flux $R^{mnp}$ cannot be the derivative of an $O(6,6)$ transformation since the derivative
has an index down. A tri-vector would be generated if we introduced a derivative with an index up, or in other words, if we doubled
the coordinates by adding dual coordinates. This is the spirit of the double torus construction of \cite{dt1,dt2}. Even though we will stick to a six-dimensional manifold, from a purely group-theoretic point of view we shall introduce a ``derivative up" $\partial^m$, T-dual to the standard derivative $\partial_m$, such that it combines with the latter to form a 12-dimensional 1-form $\partial_A$. The $\rep{220}$ representation of the fluxes is therefore obtained by
\beq \label{O66fluxes}
F_{ABC}=\partial_{[A} A_{BC]}  \ .
\eeq
where $A_{BC}=\eta_{BD} A^D{}_C$ and $A^D{}_C$ is a generic $o(6,6)$ element as introduced in (\ref{TdualgenO66}).

Out of the 220 fluxes in (\ref{O66fluxes}), half of them ($H_3$ and $\omega$) are geometric, i.e. are (standard) derivatives of elements in $\Ggeom$.
The other 110 are divided into 90 ``locally geometric" fluxes $Q$ and 20 ``locally non-geometric" fluxes $R$. The locally geometric fluxes
can be built using standard derivatives, but acting on elements of $O(d,d)$ which are not in $\Ggeom$. They are locally geometric since, as we argued
in the previous section,
non-geometric elements of $O(d,d)$ can be locally gauged away by $O(d) \times O(d)$ transformations without changing the metric and the $B_2$ field.
However, the $O(d) \times O(d)$ transformation needed to rotate the non-geometric element into a geometric one will not be single-valued if the non-geometric
element is not globally well defined.\footnote{Examples of this are given in \cite{TGG}.} On the contrary, locally non-geometric fluxes require
a non-standard derivative, and therefore cannot be the field strengths of any $O(d,d)$ gauge field.

\section{Exceptional Generalized Geometry and U-dual gauge fields}
\label{sec2}

Following \cite{E7}, in this section we incorporate also the RR fields to the geometry, similarly to what we did for $B_2$ in the previous section. With that aim we extend the structure group of the generalized tangent bundle to $E_7$.

\subsection{The $\rep{56}$ representation of the gauge parameters}
\label{sec56}

In order to geometrize the RR fields at the same time as the $B_2$ field, one needs to extend the generalized tangent bundle to one whose structure group is the full U-duality group $\E7$. The 12-dimensional generalized tangent space hence should get extended to a 56-dimensional one \cite{HullEGG,EGG}. The fundamental $\rep{56}$ representation of $\E7$ decomposes under $\Tsub \subset \E7$ as
\begin{equation}
\label{56}
\begin{aligned}
   \rep{56} & = (\rep{12},\rep{2}) + ({\rep{32'}},\rep{1})\ ,\\
   \lambda & = \left( \lambda^{Ai},\, \lambda^- \right)\ .
\end{aligned}
\end{equation}
where $i=1,2$, and a minus denotes a sum of odd forms on the internal space.\footnote{We are choosing conventions where the $\rep{56}$ contains the $\rep{32}'$ representation of $O(6,6)$, while the $\rep{32}$ appears in the $\rep{133}$ representation.
This is the appropriate choice for type IIB compactifications, while in type IIA the opposite choice is required.} The 56 degrees of freedom can be accounted for by gauge parameters. Six come from gauge transformations of the $B_2$ field, given by one-forms, and six from vectors pointing in the directions of the diffeomorphisms. Those build the fundamental ${\bf 12}$ representation of $O(6,6)$. We have to add to them gauge transformations
of the RR fields, given by odd forms in the $\rep{32'}$ representation (in this paper we will concentrate on type IIB, where the RR potentials are even, and their gauge transformations are given by odd forms). However, in order to fill out the ${\bf 56}$ representation  (or in other words, to have a closed set under U-duality), we need another 12 parameters. These are the magnetic duals in the NS sector, namely a 5-form, corresponding to gauge transformations of $B_6$ (the dual of $B_2$),
and the duals of the diffeomorphism vectors, given by elements of $T^*M \otimes \Lambda^6 T^*M$ (whose  corresponding gauge charges are the Kaluza-Klein monopoles). The 56-dimensional exceptional generalized tangent space is therefore given by
\begin{equation}
\label{eq:EGT}
   E  = TM \oplus T^*M
            \oplus \Lambda^5T^*M
            \oplus \left(T^*M\otimes\Lambda^6T^*M\right)
            \oplus \Lambda^\textrm{odd}T^*M \ .
\end{equation}
The embedding of $\GL(6) \subset \Tsub$ picks out a vector that breaks the $\rep{2}$ of $\SLR$ into $\rep{1}+\rep{1}$
such that in one direction that we will call $v^i$, the $\rep{12}$ representation of $O(6,6)$ is built out of a vector and a 1-form, while in the other (called $\omega^i$) it contains a 5-form and a 1-form tensor a volume form. This is the result of an uneven assignment of $GL(6)$ weights (see Appendix A of \cite{E7} for details). Without loss of generality one can take
\beq\label{vomega}
 v^i=(1,0) \ , \qquad \omega^i=(0,1) \ .
\eeq
Note that $v \omega\equiv\epsilon_{ij}v^j \omega^i=1$.

There are 4d gauge fields associated to each of the above gauge parameters. These come from 10d gauge fields with one space-time index.   In terms of the notation in eq.(\ref{56}), the 4d vectors are
\beq
\lambda_{\mu}=\left( (a^m{}_{\mu} +b_{m\mu} ) v^i +(\tilde b^m{}_{\mu} +\tilde k_{m\mu}  ) \, \omega^i ,  \tilde c_{\mu}^{-} \right) \ ,\label{vector}
\eeq
where $\tilde b^m{}_{\mu}\equiv\frac{1}{5!}  \epsilon^{mnopqr} b_{nopqr\mu}$ (i.e. it is the Hodge dual of the 6-form $b_6$ with one external and five internal indices), $\tilde k_{m\mu}\equiv k_{m\mu}{}^{123456}$ is the magnetic dual of the vector $a_\mu$ associated with diffeomorphisms, and $\tilde c_\mu ^- \equiv \sum_{p=2n+1} \frac{1}{p!} \epsilon^{i_1 ...i_6} c_{\mu i_1...i_p}  $ are
a sum of odd multi-vectors corresponding to the Hodge dual of the RR potentials with one space-time index.

\subsection{The $\rep{133}$ representation of the gauge fields}
\label{sec133}

The analogue of the $O(6,6)$ action on the generalized tangent space is an $\E7$ action on $E$. The adjoint $\rep{133}$ representation of $\E7$ decomposes under $\Tsub$ as
\begin{equation}
\label{133}
\begin{aligned}
   \rep{133} &= (\rep{66},\rep{1})+  (\rep{1},\rep{3})
      + (\rep{32},\rep{2})\ , \\
   A &= \left(A^{A}{}_{B}, \,  A^{i}{}_j, \, A^{+i} \right) \ .
\end{aligned}
\end{equation}
Under the $\GL(6) \subset \Tsub$ embedding, these further decompose into \cite{E7}
\begin{equation}
\begin{aligned}
   A_0 &= \left(TM\otimes T^*M\right)
            \oplus  \Lambda^2T^*M \oplus \Lambda^2TM \oplus
            \\ & \qquad \qquad
            \oplus \bbR \oplus \Lambda^6T^*M \oplus \Lambda^6TM
            \oplus \Lambda^\textrm{even}T^*M
            \oplus \Lambda^\textrm{even}TM \ .
\end{aligned}
\end{equation}
We recognize the first line to be the adjoint {\bf 66} representation of $O(6,6)$ (diffeomorphisms, b-transforms and $\beta$-transforms). The $p$-form elements on the second line correspond to $b_6$ and $c^+$ transformations, which shift the value of $B_6$ and $C^+$. Diffeomorphisms, $b_2$, $b_6$ and $c^+$ transformations form the geometric subgroup $\Ggeom \subset \E7$ used to patch the exceptional generalized tangent space (see more details in \cite{EGG,E7}). Using the notation in (\ref{133}) we embed the generators of the geometric subgroup of $\E7$ in the following way \cite{E7}
\beq \label{mugeomH}
A_{\rm{geom}}=\left( \left(\begin{array}{cc}
a & 0 \\ b_2 & -a^{T}
\end{array}  \right) , \, b_6 \, v^i v_j +\hat c_0 (v^i \omega_j+\omega^i v_j)  ,  c^+ v^i \right)
\eeq
where $c^+=c_0+c_2+c_4+c_6$ and $v_i=\epsilon_{ij} v^j$.

As in the generalized geometric case, the geometric subgroup is a semi-direct product $\Ggeom=G_{\cal A} \rtimes GL(d)$, where $G_{\cal A}$ is the nilpotent subgroup corresponding to the shift symmetries $A_p \to A_p + a_p $ for the $p$-form gauge fields $B_2, B_6$ and $C^+$. The remaining elements of $\E7$ play an analogous role to the $\beta$-transformation in the $O(6,6)$ case: they can locally be gauged away by $SU(8) \subset \E7$ transformations that do not
change the metric, dilaton, $B$ and $C$-fields (which together define an $SU(8)$ structure), but in the case where they are not globally well defined, they might carry some topologically non-trivial flux that signals a non-geometric background. Besides the bi-vector $\beta_2$, the non-geometric gauge fields are a sum of even vectors that we will call $\gamma^+$ (which are $SL(2,\mathbb{R})$-dual to the RR gauge fields), a scalar $\hat c_0$, and a six-vector $\beta_6$ that together with $b_6$ form the triplet of $SL(2,\mathbb{R})$. The full $\rep{133}$ representation arranges in the following way with respect to  the $\Tsub$ decomposition
\beq \label{muH}
A= \left( \left(\begin{array}{cc}
a & \beta_2 \\ b_2 & -a^{T}
\end{array}  \right), b_6 \, v^i v_j + \hat c_0 (v^i \omega_j+\omega^i v_j)  +\beta_6 \, \omega^i \omega_j ,  c^+v^i +  \gamma^+  \omega^i\right) \ .
\eeq

\subsection{From $E_7$ to $\Tsub$}
\label{sec:e7break}

To make contact between $E_7$ and Generalized Geometry, we have decomposed the fundamental and adjoint representations of $E_7$ into representations of $O(6,6)\times SL(2,\mathbb{R})$. In this decomposition, the $O(6,6)$ subgroup is the one of Generalized Geometry, whereas the $SL(2,\mathbb{R})$ factor corresponds to fractional linear transformations of the complex axion-dilaton which appears in heterotic string compactifications,
\begin{equation}
S_H=B_6+i e^{-2\phi} {\rm vol}_6\label{hetS}
\end{equation}
The embedding of $O(6,6)\times SL(2,\mathbb{R})$ into $E_7$ is, however, not unique. For instance, while T-duality (in the adjoint of $O(6,6)$) acts straightforwardly on eq.(\ref{muH}), type IIB S-duality seems more complicated. The latter should exchange $b_2$ and $c_2$, which are contained in different representations of $\Tsub$. This implies that S-duality is not in the $\SLR$ piece, but it is a combination of generators in both $\SLR$ and $O(6,6)$.

We can therefore select another decomposition $\Tsubb \subset \E7$ for which the $\SLR$ subgroup is the one that contains type IIB S-duality, which acts on
\begin{equation}
S_{B}=C_0 + i e^{-\phi}\label{iibS}
\end{equation}
by fractional linear transformations. In this basis, eq.(\ref{muH}) is reexpressed as,
\begin{eqnarray} \label{muB}
A|_B= \left(
\left(\begin{array}{cc} a^T & \gamma_4  \\ c_4 & -a \end{array}  \right),  c_0 \, \tilde v^i \tilde v_j+ \hat c_0 (\tilde v^i \tilde \omega_j+\tilde \omega^i \tilde v_j) + \gamma_0 \, \tilde \omega^i \tilde \omega_j  , \right. \nn \\
 \left.   (\beta_2 + \gamma_6+c_2+b_6) \, \tilde v^i + (\gamma_2+ \beta_6 + b_2 + c_6) \, \tilde \omega^i \right) \,
\end{eqnarray}
Note that type IIB S-duality now corresponds to the exchange $\tilde \omega \to \tilde v$, $\tilde v \to - \tilde \omega$. Thus, from eq.(\ref{muB}) we observe that $\gamma_2$ is the S-dual of $\beta_2$ (which is itself T-dual of $b_2$), while $\beta_6$ and $\gamma_6$ are also related by S-duality.

In what follows, we will refer to the above two decompositions as $\Tsubh$ and $\Tsubb$. To relate fields in one decomposition to fields in the other, it is useful  to assign every gauge parameter in the $\rep{56}$ representation and every gauge field in the $\rep{133}$
a weight vector, in the same spirit than e.g. \cite{ferrara}. A convenient way to write the $E_7$ roots is as vectors lying in a seven dimensional subspace of an eight dimensional vector space orthogonal to $e_7+e_8$, where $e_{i}$, $i=1,\ldots,8$, are orthonormal basis vectors. The full set of weights for the $\E7$ representations we will deal with are given in Appendix \ref{appa}. We can choose the $\SLR$ vectors $v^i$ and $\omega^i$  such that
the roots corresponding to $c^+$ are positive in the conventions of Appendix \ref{app:roots}. This
requires
\beq
v^i=-e_7+e_8 \ , \qquad \omega^i= e_7-e_8 \ .
\eeq
Given that choice, the assignment of weights is unique in order to reproduce the algebra satisfied by (\ref{muH}), given in eq.(\ref{adjac}). We summarize the weight of each gauge field in Table \ref{ta:133} below, where we have made use of the shorthand notation $\pm\equiv\pm\frac12$.

Weight vectors for the $\Tsubh$ and $\Tsubb$ decompositions of $\E7$ are related by a change of basis. Given the assignment of roots in the $\Tsubh$ basis, in order to find such change, it is enough to require that $b_2$ and $c_2$ form a doublet of $S$-duality in the new basis, and that diffeomorphisms are mapped to themselves. For convenience, we take the case on which $a\to a^T$. Calling $\tilde e_a$, $a=1,...,8$, the orthonormal basis vectors in the type IIB basis, we get the following dictionary between the two bases
\begin{eqnarray}
e_i&=& \frac14 \sum_{j=1}^6 \tilde e_j+\frac14 ( \tilde e_7- \tilde e_8)-\tilde e_i\quad  i,j=1,...,6 \label{change1} \\
e_7-e_8&=& \frac12 ( \tilde e_7- \tilde e_8)-\frac12 \sum_{j=1}^6  \tilde e_j \label{change2} \ .
\end{eqnarray}
Thus, for instance
\begin{eqnarray}
(\underline{1,0,0,0,0,0};+,-)|_H  &\to&    (\underline{-1,0,0,0,0,0};+,-)|_B \nn \\
(0,0,0,0,0,0;1,-1)|_H  &\to&    (-,-,-,-,-,-;+,-)|_B \nn
\end{eqnarray}

This change of basis dictates the form of the full $\rep{133}$ for the $\Tsubb$ decomposition, which has the structure given in eq.(\ref{muB}). We present also in Table \ref{ta:133} the assignment of weights in this type IIB basis.

\begin{table}[!ht]
\begin{center}
\begin{tabular}{|c|c|c|c|c|} \hline
  \rm{field} & $\Tsubh$ & $\Tsubb$ & $(-1)^{F_L}$& $\Omega_P$ \\ \hline
 $b_{123456}$ & $(0,0,0,0,0,0;-1,1)$ & $(+,+,+,+,+,+;-,+) $  & $+$ & $-$ \\ \hline
 $b_{12}$  &  $(1,1,0,0,0,0;0,0)$ & $(-,-,+,+,+,+;+,-)$ & $+$ & $-$ \\ \hline
 $a^1{}_2$ & $(-1,1,0,0,0,0;0,0)$ & $(1,-1,0,0,0,0;0,0)$ &  $+$ & $+$\\ \hline
 $c_0$& $(-,-,-,-,-,-;-,+)$ & $(0,0,0,0,0,0;-1,1) $ & $-$ & $-$ \\ \hline
 $c_{12}$ & $(+,+,-,-,-,-;-,+)$ & $(-,-,+,+,+,+;-,+) $ & $-$ & $+$ \\ \hline
 $c_{1234}$ & $(+,+,+,+,-,-;-,+)$ & $(0,0,0,0,1,1;0,0) $ & $-$ & $-$ \\ \hline
 $c_{123456}$ & $(+,+,+,+,+,+;-,+)$ & $(+,+,+,+,+,+;+,-)$ & $-$ & $+$ \\ \hline \hline
$\beta^{12}$  &  $ (-1,-1,0,0,0,0;0,0)$ & $(+,+,-,-,-,-;-,+)$ & $+$ & $-$ \\ \hline
$ \beta^{123456}$  & $(0,0,0,0,0,0;1,-1) $  & $(-,-,-,-,-,-;+,-)$ & $+$ & $-$ \\ \hline
$\gamma_0 $ & $(+,+,+,+,+,+;+,-)$ & $(0,0,0,0,0,0;1,-1) $ & $-$ & $-$ \\ \hline
$\gamma^{56}$  & $(+,+,+,+,-,-;+,-) $ & $(-,-,-,-,+,+;+,-)$ & $-$ & $+$ \\ \hline
$\gamma^{3456}$ & $(+,+,-,-,-,-;+,-) $ & $(-1,-1,0,0,0,0;0,0)$  & $-$ & $-$\\ \hline
$\gamma^{123456}$  & $(-,-,-,-,-,-;+,-)$ & $(-,-,-,-,-,-;-,+) $ & $-$ & $+$ \\ \hline \hline
$a^i{}_i$ & 6 Cartans & 6 Cartans & $+$ & $+$ \\ \hline
$\hat c_0$ & 1 Cartan & 1 Cartan & $+$ & $+$ \\ \hline
\end{tabular}
\caption{\small
\text{Weights of the geometric and non-geometric gauge transformations.}}\label{ta:133}
\end{center}
\end{table}

From Table \ref{ta:133} we observe that (in the $\Tsubh$ basis)
$\Ggeom$ contains all positive roots, the Cartans and a few negative roots (corresponding to $a^i{}_j, i<j$) which are minus a simple root.\footnote{In the $\Tsubb$ basis one can change conventions such that the same thing happens.} This is referred to as a parabolic subgroup.
The further subalgebra containing just the shift symmetries $b_2$, $b_6$ and $c^+$ is its unipotent radical. The geometric subgroup $G_{\rm geom}$ contains in particular the Borel subgroup of $E_7$, generated by all positive roots and Cartans. As it has been commented in the previous subsection, it is possible to locally gauge away all the transformations which lie outside the Borel subgroup by means of $SU(8)$ transformations. The 70 dimensional space of fields in the Borel subgroup (the dilaton, the metric and the $B$ and $C^+$ fields) define a $SU(8)$ structure on the generalized tangent bundle and parameterize the coset $E_7/SU(8)$. This is also the space of scalar fields of the resulting 4d effective supergravity ($7$ fundamental scalars corresponding to the Cartans, $48$ axions and $15$ diffeomorphisms corresponding to the positive roots, which transform non-linearly under $E_7/SU(8)$ \cite{andrianopoli1,andrianopoli2}).

From the point of view of string theory compactifications the decomposition of the $E_7$ structure into a $O(6,6)\times SL(2,\mathbb{R})$ subgroup is the result of orientifolding the theory.\footnote{Note that we are making a rather general use of the term `orientifolding', referring also to the non-perturbative description of an orientifold. In this general sense, the heterotic string would be for instance considered an orientifold of type IIB String Theory through the duality chain type IIB $\to$ type I $\to$ heterotic.\label{footori}} Whereas in general this introduces $n$ extra vector multiplets in the theory, as required by anomaly cancellation, and the global symmetry group is enhanced to $O(6,6+n)\times SL(2,\mathbb{R})$, here we choose to focus on the set of (untwisted) states which come from truncation of the parent unorientifolded theory. This set is closed under a subgroup $O(6,6)\times SL(2,\mathbb{R})\subset O(6,6+n)\times SL(2,\mathbb{R})$. The simplest examples are toroidal compactifications, where the $E_7$ structure group corresponds to the U-duality group of the resulting 4d $\mathcal{N}=8$ supergravity. After orientifolding only half of the supersymmetries are preserved, and $E_7$ is broken to the U-duality group of the resulting 4d $\mathcal{N}=4$ supergravity, $\Tsub$. In this context, the change of basis vectors in eqs.(\ref{change1})-(\ref{change2}) corresponds to T-dualizing along all the coordinates of the internal $T^6$ and applying type I - $SO(32)$ heterotic duality, thus dualizing from a type IIB compactification with O3-planes to a heterotic compactification. By comparing the assignment of weights for the two basis in Table \ref{ta:133} (or alternatively eqs.(\ref{muH}) and (\ref{muB})) we can see how gauge transformations are mapped under this type IIB - heterotic duality. For instance, the heterotic $b_2$ is mapped to the type IIB $c_4$, whereas the heterotic $b_6$ is mapped to the type IIB $c_0$. We will see more details on toroidal compactifications with general fluxes in Section \ref{sectori}.

One may easily check that states transforming in spinorial representations of $O(6,6)$ (e.g. $\mathbf{32}$, $\mathbf{32}'$, $\mathbf{352}$, etc.) have negative parity under the orientifold action, whereas the remaining states have positive parity. In particular, note that the gauge fields
surviving the orientifold projection are precisely those in the
adjoint of $O(6,6)\times SL(2)$, for both decompositions.
In terms of weight vectors, we can therefore introduce an operator $P$ such that it acts on a given state $k$ as
\begin{equation}
P(k)=(-1)^{k\cdot u}k \ , \quad \textrm{with} \quad   u = (1, 1, 1, 1, 1, 1; 0, 0)
\end{equation}
This operator is identified in heterotic compactifications with the space-time fermionic number for left-movers $(-1)^{F_L}$, whereas in the type IIB basis corresponds to the orientifold action $\Omega_P(-1)^{F_L}\sigma$, where $\Omega_P$ is the worldsheet parity and $\sigma$ an orientifold involution which reverses all coordinates of $T^6$. Acting with $P$ on the states in Table \ref{ta:133} we see indeed that $\hat c_0$, $a$, $b_2$, $b_6$, $\beta_2$ and $\beta_6$ are kept in the $\Tsubh$ basis, whereas $a$, $\hat c_0$, $c_0$, $c_4$, $\gamma_0$ and $\gamma_4$ are kept in the $\Tsubb$ basis. We have stated in Table \ref{ta:133} the parity for the different fields under $(-1)^{F_L}$ and $\Omega_P$ obtained in this way.

\section{U-dual field strengths}
\label{sec2b}

The discussion of field strengths associated to U-duality covariant gauge potentials follows closely the one for T-dual field strengths in Section \ref{sec1}. These were obtained by tensoring the adjoint representation of $O(6,6)$, containing the T-duality covariant gauge potentials, with the vector representation, containing an extension of the standard derivative which also accounts for states with non-zero winding. Field strengths were obtained by projecting to the antisymmetric part of the tensor product. In the present context, the same procedure amounts to tensoring the $\mathbf{56}$ and $\mathbf{133}$ representations of $E_7$ that we have already introduced. In terms of irreducible representations the tensor product decomposes as (see e.g. \cite{slansky}),
\begin{equation}
\mathbf{56}\times\mathbf{133}=\mathbf{56}+\mathbf{912}+\mathbf{6480} \ .
\end{equation}
Field strengths are then identified with the $\mathbf{912}$ representation of $E_7$ \cite{dewit}.

The $O(6,6)$ tensorial structure of the field strengths can be determined from their weights in the $\Tsubh$-basis, as the $O(6,6)$ factor is identified in this basis with the structure group of the generalized tangent bundle. Writing a generic element in the  $\rep{912}$ as
\begin{equation}
\label{912}
\begin{aligned}
   \rep{912} &= (\rep{12},\rep{2})+  (\rep{32'},\rep{3})
      + (\rep{352},\rep{1}) + (\rep{220},\rep{2}) \ , \\
   f  &= \left(f^{Ai}, \,  f^{i}{}_{j}{}^-, \, f^{A+}, \, f^{ABCi}  \right) \ .
\end{aligned}
\end{equation}
we have in the $\Tsubh$ basis\footnote{Our notation for some of the fluxes is slightly different from that of \cite{acfi}. In particular, $H'\leftrightarrow F'$ and $P'\leftrightarrow Q'$ are exchanged. We find this notation more suited to the $(-1)^{F_L}$ charges of the fluxes. Moreover, $Q\to -Q$ and $\omega \to -\omega$ with respect to \cite{acfi}.}
\beq
\begin{aligned}\label{fluxes332220}
f^{Ai}&=v^i (\bom_a + \bQ^
a)  + \omega^i (\bQ'_a + \bom'^
a) \epsilon^{123456} \ , \\
f^{i}{}_{j}{}^-&=F^- v^i v_j + \bP^- \epsilon^{123456} (v^i \omega_j+\omega^i v_j) +F'^{-} (\epsilon^{123456})^2 \omega^i \omega_j  \ , \\
f^{A+}&=\left(\frac{1}{2} P_{m,i_1i_2} + \frac{1}{4!} P_{m,i_1i_2i_3i_4} + \frac{1}{6! }\hat F_m \epsilon_{i_1...i_6} \right) \epsilon^{i_1...i_6} \\
& \quad \left( \hat F'^m +
 \frac{1}{2} P'^m_{i_1 i_2} +  \frac{1}{4!} P'^m_{i_1 i_2i_3 i_4}  \right)\epsilon^{i_1,...,i_6}\\
f^{ABCi}&= v^i \left(H_{abc}+\omega_{ab}^c+Q^{ab}_c+R^{abc}\right) + \omega^i \left(\tilde H_{abc}+Q'^{a}_{bc}  +\omega'^{ab}_c +H'^{abc}  \right) \epsilon^{123456} \ ,
\end{aligned}
\eeq
where the traceless condition on the $\rep{352}$ representation (which corresponds to  a traceless vector-spinor) is encoded in the absence of a 0-form in the first line, the absence of a 6-form in the second line, and the extra condition $P'^m_{m i_2i_3 i_4} =0$ that has to be imposed on this flux. Note that
as defined in eq.(\ref{fluxes332220}), $f^{A+}$ does not satisfy $\Gamma_A f^{A+}=0$, but it has the same number of degrees of freedom of a traceless vector-spinor. We summarize in Table \ref{tabla352} of Appendix \ref{912weight} the assignment of $E_7$ weights for the field strengths in both basis and their parity under $(-1)^{F_L}$ and $\Omega_P$.\footnote{Depending on the context, we represent the tensor structure of the fluxes in slightly different ways by making use of the 6d antisymmetric tensor. Thus, for instance,
\begin{align*}
&\hat F'^{1,123456}\equiv\hat F'^{1}\epsilon^{123456}\\
&P_{1}^{2345}\equiv P_{1,16}\epsilon^{123456}, \ \ \textrm{etc.}
\end{align*}
The notation has been chosen in such a way that there is not possible ambiguity.\label{footnot}}

To shed light on the 10d supergravity uplift of these field strengths we can compute their explicit expression in terms of derivatives of the fields in Table \ref{ta:133}, similarly to what we did in eqs.(\ref{tfield}) for T-dual field strengths. In the language of representation theory, this is equivalent to computing the Clebsh-Gordan coefficients for the $\mathbf{912}$ representation expressed in terms of elements of the $\mathbf{56}$ and $\mathbf{133}$.

With that aim, we take the highest weight of the $\mathbf{912}$ representation
expressed as a linear combination of weights belonging to the tensor product $\mathbf{56}\times \mathbf{133}$,
\begin{align}
(+,+,+,+,+,-;-1,1)&=\frac{1}{\sqrt{7}}\left[(1,0,0,0,0,0;-,+)\times(-,+,+,+,+,-;-,+)\ \right.\label{highest}\\
&-\ (0,1,0,0,0,0;-,+)\times(+,-,+,+,+,-;-,+)\ \nonumber\\
&+\ (0,0,1,0,0,0;-,+)\times(+,+,-,+,+,-;-,+)\ \nonumber\\
&-\ (0,0,0,1,0,0;-,+)\times(+,+,+,-,+,-;-,+)\ \nonumber\\
&+\ (0,0,0,0,1,0;-,+)\times(+,+,+,+,-,-;-,+)\ \nonumber\\
&-\ (0,0,0,0,0,-1;-,+)\times(+,+,+,+,+,+;-,+)\ \nonumber\\
&\left.+\ (+,+,+,+,+,-;0,0)\times (0,0,0,0,0,0;-1,1)\right]\nonumber
\end{align}
The numerical coefficients have been determined in such as way that the r.h.s. of the equation vanishes when acted with any positive root of $E_7$, as corresponds to the highest weight of a representation. We refer the reader to Appendix \ref{appa} for further details on the $E_7$ algebra and root system.

In terms of the elements in Tables \ref{ta:133} and \ref{tabla352}, eq.(\ref{highest}) reads,
\begin{equation}
\label{f5}
F_{12345}=5\partial_{[1}c_{2345]}-\partial^6c_{123456}+\hat\lambda^6 b_{123456}
\end{equation}
where, for convenience, we have introduced a generalized exterior derivative in the $\rep{56}$ representation
\beq D\equiv \left((\partial_m +\partial^m)v^i+ (\tilde\partial_m +\tilde\partial^m)\omega^i, \hat \lambda^- \right)
\label{deriv56}\eeq
The assignment of weights is presented in Table \ref{ta:56}.
\begin{table}[!ht]
\begin{center}
\begin{tabular}{|c|c|c|} \hline
 $D$&$\Tsubh $ & $\Tsubb$   \\ \hline
$\partial_1$ & $(1,0,0,0,0,0;-,+)$  &$ (-,+,+,+,+,+;0,0)$ \\ \hline
$\partial^1$ & $(-1,0,0,0,0,0;-,+)$  &$  (1,0,0,0,0,0;-,+)$ \\ \hline
$\tilde \partial^{1} \epsilon $ & $(-1,0,0,0,0,0;+,-)$  &$  (+,-,-,-,-,-;0,0)$ \\ \hline
$\tilde \partial_1 \epsilon$ & $(1,0,0,0,0,0;+,-)$  &$ (-1,0,0,0,0,0;+,-)$ \\ \hline
$\hat \lambda^6$& $(+,+,+,+,+,-;0,0)$ & $(0,0,0,0,0,1;+,-)$\\ \hline
$\hat \lambda^{456}$& $(+,+,+,-,-,-;0,0)$ & $(-,-,-,+,+,+;0,0)$\\ \hline
$\hat \lambda^{23456}$& $(+,-,-,-,-,-;0,0)$ & $(-1,0,0,0,0,0;-,+)$\\ \hline
\end{tabular}
\caption{\small
Assignment of weights in the 56 representation.
The symbol $\epsilon$ means $\epsilon^{123456}$, i.e. an inverse volume factor.}\label{ta:56}
\end{center}
\end{table}

Note that the first term in the r.h.s. of eq.(\ref{f5}) corresponds to the standard definition of the 5-form RR field strength, whereas the second and third terms correspond to locally non-geometric contributions.

Acting on eq.(\ref{highest}) with the generators associated to negative roots $E_{-\alpha_i}$, $i=1,\ldots, 7$ (c.f. eqs.(\ref{alpha}) in the Appendix) we can build similar relations for the other elements in the $\mathbf{912}$. For instance, acting on eq.(\ref{highest}) with $E_{-\alpha_6}E_{-\alpha_4}E_{-\alpha_5}$ leads to,
\begin{equation}
F_{123}=3\partial_{[1}c_{23]}+\partial^ic_{123i}+\hat\lambda^{456} b_{123456}
\end{equation}

The procedure can be systematized with the aid of the computer.\footnote{There are some subtleties that have to be taken into account, however. In particular, notice that there can be independent sequences of negative roots which result in the same weight of $\mathbf{912}$. This is the origin of the multiple copies of the same weight appearing in the $(\mathbf{220},\mathbf{2})$ and $(\mathbf{12},\mathbf{2})$, or in the $(\mathbf{352},\mathbf{1})$ and $(\mathbf{32}',\mathbf{3})$ (c.f. Appendix \ref{appa}). In order to disentangle weights appearing in various representations of $O(6,6)\times SL(2,\mathbb{R})$ it is important to stress that in the conventions of Appendix \ref{appa}, $E_{-\alpha_i}$, $i=1,\ldots, 6$ is also a basis of negative roots of $O(6,6)$, whereas the negative root of $SL(2,\mathbb{R})$ is given by the combination
\begin{equation*}
[E_{-\alpha_7},[\ldots[E_{-\alpha_3},[E_{-\alpha_2},E_{-\alpha_1}]]\ldots]]
\end{equation*}
where the sequence of subindices is $7,6,4,5,3,4,6,7,2,3,4,6,5,4,3,2,1$.} In terms of eqs.(\ref{56}) (with $\lambda \to D$), (\ref{133}) and (\ref{912}) we get,
\begin{align}
f^i{}_j{}^-&=-\epsilon_{jk}D^{M(i}\Gamma_MA^{+k)}+D^-A^i{}_j\label{dec1}\\
f^{MNPi}&=3D^{i[M}A^{NP]}+\langle D^-,\Gamma^{MNP}A^{+i} \rangle \\
f^{Mi}&=2D^{i[N}A^{M]P} \eta_{NP}+2 D^{Mj} A^i{}_j + \langle D^-,\Gamma^MA^{+i}\rangle\\
f^{M+}&=\epsilon_{ij} D^{Mi}A^{+j}-\frac{1}{11} \epsilon_{ij} D^{Ni} \Gamma_N{}^M A^{+j}-A^M{}_N \Gamma^N D^- + \frac{1}{10} A_{NP} \Gamma^{MNP} D^-\label{dec2}
\end{align}
where $\Gamma^M$ are the $O(6,6)$ Gamma matrices, which act on forms by\footnote{$\iota_m$ means a contraction along $m$, which acts like $\iota_m \frac{1}{p!} A_{i_1 ... i_p} dx^{i_1} \wedge dx^{i_p} = \frac{1}{(p-1)!} A_{mi_2 ... i_p} dx^{i_2} \wedge dx^{i_p}$.}
\beq \label{Gamma}
  \Gamma^A \leftrightarrow (dx^m \wedge , \iota_m )
  \eeq
and the bracket denotes the Mukai pairing defined by
\begin{equation}
\label{mukai}
   \mukai{\psi}{\chi}
     = \sum_p (-)^{[(p+1)/2]} \psi_p \wedge \chi_{6-p}
     \ .
\end{equation}

Equations (\ref{dec1})-(\ref{dec2}) encode the expression of all the field strengths contained in the $\rep{912}$ representation of $E_7$, expressed in terms of generalized derivatives of the gauge potentials in the $\rep{133}$ representation. We can recast them in a more standard form by making use of eqs.(\ref{muH}), (\ref{fluxes332220}) and (\ref{deriv56}).
For simplicity we only present here explicitly the field strengths which involve standard derivatives of the potentials in Table \ref{ta:133} (i.e. we set to zero all the exotic derivatives in eq.(\ref{deriv56})), and which therefore admit an uplift to locally geometric solutions of 10d type IIB supergravity. We can organize them as follows:
\begin{enumerate}
\item \underline{NSNS fluxes:}
\begin{equation}
H_{ijk}=3\partial_{[i}b_{jk]}\label{h1}
\end{equation}
\item \underline{RR fluxes:}
\begin{equation}
F_{ijklm}=5\partial_{[i}c_{jklm]}\ , \qquad F_{ijk}=3\partial_{[i}c_{jk]}\ , \qquad F_{i}=\partial_{i}c_{0}
\end{equation}
\item \underline{Metric fluxes:}
\begin{equation}
\omega_{ij}^k=2\partial_{[i}a^{k}{}_{j]}\ , \qquad \bom_{p}=\partial_{i}a^i{}_{p}\label{metric}
\end{equation}
\item \underline{$\beta$-deformations (NSNS):}
\begin{equation}
Q^{ij}_k=\partial_{k}\beta^{ij}\ , \qquad \bQ^
{p}=\partial_{i}\beta^{ip}\ , \qquad \bQ'^{123456}_i=-\partial_i\beta^{123456}\label{qs}
\end{equation}
\item \underline{$\gamma$-deformations (RR):}
\begin{align}
&P_i^{jk}=\partial_i\gamma^{jk}\ , \qquad P_i^{jklm}=\partial_i\gamma^{jklm}\ , \qquad P'^{i,jklm}=\partial_q\gamma^{ijklmq} \ , \qquad \hat F_i=\partial_i\gamma_0\ , \nonumber\\
&\qquad \qquad \bP^
k=\partial_i\gamma^{ik}\ , \qquad \bP^
{klm}=\partial_i\gamma^{iklm}\ , \qquad \bP^
{klmrs}=\partial_i\gamma^{iklmrs}\label{ps}
\end{align}
\end{enumerate}
Any vacuum of the 4d theory involving only these backgrounds should admit a consistent, locally geometric, description in terms of 10d type IIB supergravity. Indeed, the uplift to 10d is almost automatic for NSNS, RR and metric fluxes, as these field strengths are in one to one correspondence with field strengths of type IIB supergravity. From the algebraic point of view they correspond to derivatives of elements in $G_{\rm geom}$. On the other hand, elements which lie outside the Borel subalgebra ($\beta$- and $\gamma$-deformations) in general require also local $SU(8)$ transformations to be described as a locally geometric 10d background.

Backgrounds of 10d supergravity which involve $\beta$-deformations have been considered in the recent literature \cite{luninmaldacena,minasian}, resulting important in the context of the AdS/CFT correspondence for understanding some of the marginal deformations of $\mathcal{N}=4$ super Yang-Mills \cite{ls}. Backgrounds of 10d supergravity involving $\gamma$-deformations, on the other hand, have been much less studied (see, however, \cite{acr} for some partial results). As it has been commented, $\gamma$-deformations provide the RR counterpart of the $\beta$-deformation.

The orientifold projection selects field strengths which lie in the $(\mathbf{220},\mathbf{2})+(\mathbf{12},\mathbf{2})$ representation. These can be read directly from Table \ref{tabla352}, for both type IIB with O3-planes ($\Tsubh$ basis) and heterotic string compactifications ($\Tsubh$ basis). We summarize the surviving set of fields in Tables \ref{220S} and \ref{12S}, where we write in parenthesis the components which do not admit a locally geometric interpretation in terms of 10d supergravity.

For type IIB orientifold compactifications, $\gamma$-deformations provide the complexification of the $\beta$ parameter of the deformed 4d super Yang-Mills theory in the worldvolume of D3-branes,
\begin{equation}
\beta_2-S_B\gamma_2 \label{complexbeta}
\end{equation}
where $S_B$ is the type IIB complex dilaton defined in eq.(\ref{iibS}). As we will see in Section \ref{sec4}, $\mathcal{N}=1$ supersymmetry equations require the above combination to be an anti-holomorphic $(0,2)$ complex bi-vector. We postpone further comments on this type of backgrounds to that Section.

It is also interesting to stress that some of the field strengths in eqs.(\ref{metric})-(\ref{ps}) transform in the $(\mathbf{12},\mathbf{2})$ representation, either with respect to the $\Tsubh$ basis ($\bom_{i}$, $\bQ^i$ and $\bQ'^{123456}_i$) or to the $\Tsubb$ basis ($\bQ^i$, $\bP^i$, $\bP^{ijkrs}$ and $\bQ'^{123456}_i$). Hence, according to Table \ref{12S}, most of the field strengths in the $(\mathbf{12},\mathbf{2})$ representation admit a priori an uplift to 10d locally geometric type IIB or heterotic supergravity backgrounds. These are related to non-traceless metric fluxes, $\beta$- and $\gamma$-deformations. These field strengths will turn out particularly relevant in the context of 4d $\mathcal{N}=4$ gauged supergravity, as it will become manifest in next Section.

\begin{table}[h!]
\begin{center}
\begin{tabular}{|c|c|c|}
\hline
weight & IIB with D3/D7 & heterotic \\
\hline
$(1,1,1,0,0,0\ ;-\frac{1}{2},\frac{1}{2})$ & $F_{456}$&   $H_{123}$ \\
\hline
$(1,1,-1,0,0,0\ ;-\frac{1}{2},\frac12)$ & $Q^{12}_3$  & $\omega^3_{12}$ \\
\hline
$ 5\times (1,0,0,0,0,0\ ;-\frac{1}{2},\frac12)$ & $Q^{1m}_m$ & $\omega^m_{1m}$ \\
\hline
$(1,-1,-1,0,0,0\ ;-\frac{1}{2},\frac12)$ & $(P'^{1,1456})$ & $Q^{23}_1$ \\
\hline
$5\times (-1,0,0,0,0,0\ ;-\frac{1}{2},\frac12)$ & $P'^{2,3456}$&  $Q^{1m}_m$ \\
\hline
$(-1,-1,-1,0,0,0\ ;-\frac{1}{2},\frac12)$ & $(H'^{456,123456})$&  $(R^{123})$ \\
\hline \hline
$(1,1,1,0,0,0\ ;\frac{1}{2},-\frac{1}{2})$ & $H_{456}$& $(\tilde H_{123})$ \\
\hline
$(1,1,-1,0,0,0\ ;\frac{1}{2},-\frac12)$ & $P^{12}_3$&  $(Q'^{3,3456})$ \\
\hline
$5\times (1,0,0,0,0,0\ ;\frac{1}{2},-\frac12)$ & $P^{1m}_m$&  $(Q'^{[2,3456]})$ \\
\hline
$(1,-1,-1,0,0,0\ ;\frac{1}{2},-\frac12)$ & $(Q'^{1,1456})$ & $(\omega'^{23456,23})$ \\
\hline
$5\times (-1,0,0,0,0,0\ ;\frac{1}{2},-\frac12)$ & $(Q'^{[2,3456]})$ & $(\omega'^m_{1m})$ \\
\hline
$(-1,-1,-1,0,0,0\ ;\frac{1}{2},-\frac12)$ & $(F'^{456,123456})$&  $(H'^{123,123456})$ \\
\hline
\end{tabular}
\caption{\small
Field strengths transforming in the $(\mathbf{220},\mathbf{2})$ representation of $O(6,6)\times SL(2,\mathbb{R})$. Components between parenthesis do not admit a 10d locally geometric description.}\label{220S}
\end{center}
\end{table}

\begin{table}[h!]
\begin{center}
\begin{tabular}{|c|c|c|}
\hline
weight & IIB with D3/D7 & heterotic \\
\hline
$(1,0,0,0,0,0\ ;-\frac{1}{2},\frac{1}{2})$ & $\bQ^
1$ & $\bom_{1}$ \\
\hline
$(-1,0,0,0,0,0\ ;-\frac{1}{2},\frac12)$ & $\bP^{23456}$ & $\bQ^
1$  \\
\hline \hline
$(1,0,0,0,0,0\ ;\frac{1}{2},-\frac{1}{2})$ & $\bP^
1$ & $\bQ'^{123456}_1$ \\
\hline
$(-1,0,0,0,0,0\ ;\frac{1}{2},-\frac12)$ & $\bQ'^{123456}_1$ &$(\bom'^
1)$ \\
\hline
\end{tabular}
\caption{\small
Field strengths transforming in the $(\mathbf{12},\mathbf{2})$ representation of $O(6,6)\times SL(2,\mathbb{R})$. Components between parenthesis do not admit a 10d locally geometric description.}\label{12S}
\end{center}
\end{table}

\section{Toroidal compactifications and dual fluxes}
\label{sectori}

Up to here we have been discussing various aspects related to the local $E_7$ structure of the generalized tangent bundle, or the $\Tsub$ subgroup which survives after taking the orientifold projection. We have in particular introduced a set of gauge parameters, gauge fields and field strengths which are covariant under the structure group.

In the present (and forthcoming) Section we consider the inclusion of topologically non-trivial fluxes for these field strengths. The global structure of the tangent bundle therefore becomes relevant also. In that regard, here we consider the simplest possible situation, namely compactifications on manifolds of trivial structure (tori, or ``twisted tori"). Since tori are parallelizable manifolds, global and local structure groups in this case coincide. The more interesting case of (not necessarily toroidal) compactifications with global $SU(3)$ structure will be treated in Section \ref{sec4}.

The $E_7$ structure group of the generalized tangent bundle becomes in toroidal compactifications  also the global symmetry group of the resulting effective 4d $\mathcal{N}=8$ gauged supergravity. The $\rep{56}$ representation discussed in Section \ref{sec56} relates to the 56 vector fields of $\mathcal{N}=8$ supergravity (in its electric-magnetic covariant formulation), whereas the $\rep{912}$ representation of field strengths, discussed in Section \ref{sec2b}, is nothing but the embedding tensor of the gauged version \cite{gauged8}, thus establishing a precise dictionary between fluxes and gaugings.

After taking the orientifold projection (see footnote \ref{footori}), the structure group is reduced to $\Tsub$, accordingly to the discussion in Section \ref{sec:e7break}. This is also the global symmetry group of the resulting 4d $\mathcal{N}=4$ gauged supergravity. In this case, vector fields in the 4d theory are grouped in the $(\rep{12},\rep{2})$ representation of $\Tsub$, whereas field strengths arrange in the $(\rep{220},\rep{2})+(\rep{12},\rep{2})$ representations (c.f. Tables \ref{220S} and \ref{12S}) and correspond to the embedding tensor of 4d $\mathcal{N}=4$ gauged supergravity \cite{sw}.

In this context, it results particularly interesting the observation found in Section \ref{sec2b} that all gaugings in the $(\rep{12},\rep{2})$ representation admit an uplift to locally geometric solutions of 10d supergravity. From the 4d point of view, some of these gaugings are known to be related to twists by an axionic rescaling symmetry \cite{vzminimal}. Their higher dimensional origin, however, has been a more obscure and longstanding problem. Some of them have been identified as arising on particular Scherk-Schwarz reductions of 10d heterotic supergravity \cite{marios}. This is consistent with Table \ref{12S}, where we observe that gaugings in the first row admit an uplift to heterotic compactifications with non-traceless metric fluxes. Also, more recently it has been shown that some of these gaugings admit an uplift to type IIA orientifold compactifications with dilaton fluxes \cite{Dall'Agata:2009gv}. In this regard, the results of Section \ref{sec2b} reveal that all gaugings in the $(\rep{12},\rep{2})$ representation can be also understood as originating from type IIB orientifold compactifications with non-traceless $\beta$- and $\gamma$-deformations.

\subsection{U-dual gauge algebra and constraints}
\label{sec3}

The different fluxes, grouped in the ${\bf 912}$ representation of the U-duality group $E_7$, are expected to obey diverse constraints.  In fact, from the 10d point of view, Bianchi identities as well as tadpole cancelation equations  must be satisfied.

From a 4d perspective  fluxes are associated to the structure constants of the gauge algebra satisfied by the vectors of the effective $\mathcal{N}=8$ gauged supergravity, given in eq.(\ref{vector}). Formally, we expect to have an algebra
\begin{equation}
 [{\cal X}_{\cal A},{\cal X}_{\cal B}] =F_{\cal A \cal B}^ {\cal P} \, {\cal X}_{\cal P}
 \label{algebra}
\end{equation}
where  ${\cal X}_{\cal A}$ are the gauge generators and $F_{\cal A \cal B}^ {\cal P} $ encode the fluxes. Here ${\cal A}$, ${\cal B}$, etc. are indices of the $\rep{56}$ representation of $E_7$ (see Section \ref{sec56}). If Jacobi identities are satisfied, then fluxes are constrained to obey
\begin{equation}
F_{[\cal A\cal B}^ {\cal P}F_{\cal C]\cal D}^ {\cal L}=0
\label{extra}
\end{equation}
In \cite{acr} it was shown, for a particular type IIB orientifold,  that these Jacobi identities are actually identified with the
Bianchi identities and tadpole equations of the higher dimensional theory. Therefore, once the dictionary between structure constants and fluxes is known we should be able to read the flux constraints directly from the Jacobi identities of the 4d gauge algebra.

In a generic situation (see for instance \cite{samt}), the function $F_{\cal A \cal B}^ {\cal P}$ expressed in terms of the fluxes is not antisymmetric in the subindices ${\cal A}, {\cal B}$ and therefore does not define a consistent algebra.\footnote{In the general situation where $F_{(\cal A \cal B)}^ {\cal P}\neq 0$, in order to covariantize the Lagrangian 4d gauge bosons have to be supplemented with the 2-forms which result from dualizing the scalars of the theory \cite{magnetic}.}  In order to ensure antisymmetry of the commutators, extra constraints must be imposed. Namely,    the symmetric part of the structure constants, contracted with  generators must vanish,
$F_{(\cal A \cal B)}^ {\cal P} \, {\cal X}_{\cal P}=0$. Jacobi identities are therefore satisfied when a contraction with a third generator is considered. In terms of structure constants this new condition reads
\begin{equation}
F_{\cal A\cal B}^ {\cal P}F_{\cal P\cal C}^ {\cal L}+F_{\cal B\cal A}^ {\cal P}F_{\cal P\cal C}^ {\cal L}=0
\label{antisymm}
\end{equation}
In what follows we sketch a possible way to obtain the expression of the structure constants in terms of fluxes.
The idea is to follow similar steps as in Ref.\cite{acr}.
In other words, by starting with a previously known sector of the algebra (\ref{algebra}), we can apply $E_7$ generators to it in order to construct the entire algebra.

By looking at both sides of the gauge algebra (\ref{algebra}), we see that in terms of irreducible representations of $E_7$, the left hand side leads to
\begin{eqnarray}
 {\bf 56}\times{\bf 56}&=& ( {\bf 1}+ {\bf 1539} )_A + ( {\bf 133 }+ {\bf 1463 } )_S
\end{eqnarray}
where we have made explicit the distinction between symmetric and antisymmetric representations. Similarly, the right hand side decomposes as
\begin{eqnarray}
	{\bf 912} \times {\bf 56}&=& {\bf 1539}+ {\bf 133}+ {\bf 40755}+ {\bf 8645 }
\end{eqnarray}
Thus,  in order for both sides to match,  only products of states belonging to the ${\bf 1539}$ antisymmetric representation must be kept.

We start considering the sector of the algebra invariant under an $\Tsub \subset E_7$ subgroup. It is possible to check that the ${\bf 1539}$ representation of $E_7$ decomposes as
\begin{eqnarray}
  {\bf 1539}=( {\bf 77}, {\bf 1})+( {\bf 32},  {\bf 2})+( {\bf 66}, {\bf  3})+( {\bf 495}, {\bf  1})+( {\bf 352},  {\bf  2})+( {\bf 1}, {\bf  1})
\end{eqnarray}
Similarly, the generators of the algebra are decomposed accordingly to eq.(\ref{56}),
\begin{equation}
{\cal X}_{\cal A}= {\cal X}_{Ai}\oplus {\cal X}'_{ S^-}
\end{equation}
where $S^-$ is a weight belonging to the $\rep{32'}$ representation. Alternatively, ${\cal X}'_{ S^-}$ can be expressed in terms of a set of multi-vectors, ${\cal X}'_{ S^-}=X'_m\oplus X'_{mno}\oplus X'_{mnopq}$.

The general structure of the algebra, written in terms of  $\Tsub$ content hence reads
\begin{eqnarray}
{[{\cal X}_{ A i} \oplus {\cal X}_{ S^ -}, {\cal X}_{ B j}\oplus {\cal X}'_{ S'^ -}]}&=&
F_{\cal A\cal B}^ {Pk} {\cal X}_{Pk}+F_{\cal A \cal B}^ { S''^-} {\cal X}'_{S''^-}
\end{eqnarray}
where ${\cal A}=( A i, S^-)$ and ${\cal B}=( B j, S'^-)$. In order to start with,  we choose the   subsector given by the fluxes transforming in the $({\bf 12},  {\bf 2})+( {\bf 220},{\bf 2})$ representation of $\Tsub$
\begin{eqnarray}
{[{\cal X}_{ A i} , {\cal X}_{ B j}]}&=&
F_{ A i B j}^ {Pk} {\cal X}_{Pk}\label{redalg}
\end{eqnarray}
In this case the  explicit expression for structure constants is known to be \cite{sw}
\begin{eqnarray}
F_{Ai Bj}^ {k P}&=&  \delta_{j}^{k}f_{AB}{}^P{}_i
+ \frac12\left(\delta_{j}^{k} {f}_{Bi} \delta_{A}^{P}-\delta_{j}^{k} {f}_{Ai} \delta_{B}^{P}
+ \epsilon_{ij}    {f}_{(A|l} \epsilon^{l k}\delta_{B)}^{P}
- \eta_{AB}  {f}_{[i}^ P\delta_{j]}^{k}\right)
\label{o6sc}
\end{eqnarray}
with ${f}^{Ai}$ and $f_{ABCi}$ given in eq.(\ref{912}), while the antisymmetry constraint is
\begin{equation}
 {\epsilon}^{ij}f_{AB}{}^P{}_if_{PDLj}=0
\label{extra1}
\end{equation}

Notice that performing a  similar analysis as above  in terms of representations,  the left member of eq.(\ref{redalg}) leads to
\begin{eqnarray}
 ({\bf 12}, {\bf 2})\times({\bf 12}, {\bf 2})_{A}=( {\bf 66}, {\bf  3})+( {\bf 1}, {\bf  1})+( {\bf 77}, {\bf 1})
 \end{eqnarray}
 while the right hand side contains the structure constants given in eq.(\ref{o6sc}) times gauge generators in the $({\bf 12}, {\bf 2})$ representation, namely
  $[(  {\bf 220}, {\bf 2})+( {\bf 12}, {\bf 2} )]_{ {\bf 912}}\times( {\bf 12}, {\bf 2})$. Inspection of the structure constants indicates that  $( {\bf 66}, {\bf  3})$ comes from products with $ ( {\bf 220}, {\bf 2})$ fluxes while the rest comes from products with $( {\bf 12}, {\bf 2} ) $ fluxes.

For the sake of clarity let us present a concrete example and choose $i=1$, $j=2$ and $A=1+6$, $B=2+6$.
Then, in terms of type IIB fluxes, we read from eq.(\ref{redalg})
\begin{eqnarray}
{[\hat X^{1} , X^{2}]}&=&
Q^{12}_ p \hat X^{m} +{\tilde F}^{12m} \hat X_{p}
-P^{12}_ p X^{m} +{\tilde H}^{12m} X_{m} \label{alg1}\\
&+&  \bQ^
1 \hat X^2+     2\bQ^
2 \hat X^{1} -2\bP^
{1} X^{2}- \bP^{2} X^1\nn
\end{eqnarray}
where, in order to avoid confusion with the notation we have defined
\begin{equation}
\hat X^m\equiv {\cal X}_{(m+6)1}\ , \qquad
X^m\equiv {\cal X}_{(m+6)2}\ , \qquad m=1,\ldots ,6
\end{equation}
to denote the two elements of the $SL(2)$ doublet in $( {\bf 12},  {\bf 2})$,  with weights
$(0,1,0,0,0,0,\mp,\pm)$ (where $1$ is in the $m$-th entry). The first row of (\ref{alg1}) contains fluxes in the $( {\bf 220},{\bf 2})$ representation, summarized in Table \ref{220S}, while the second row corresponds to fluxes in the $( {\bf 12},  {\bf 2})$, presented in Table \ref{12S}.

As mentioned, by applying $E_7$ generators the full algebra can be reconstructed. For instance, by acting with $E_{\alpha_7}$, the generator corresponding to the positive simple root $\alpha_{7}  \equiv   (-,-,-,-,-,-;-,+)$,  we obtain the commutator between a vector in the $( {\bf 12},  {\bf 2})$ and a spinor in the $({\bf 32}',{\bf 1})$. Hence, acting in the left hand side of (\ref{alg1}) and taking into account that $ E_{\alpha_{7} } \hat X^{m}  \rightarrow  X'^{m}$, we observe that $ (1\otimes E_{\alpha_7} +  E_{\alpha_7}\otimes 1)
{[\hat X^{1} , X^{2}]}\rightarrow  {[X'^{1} , X^{2}]}$. Here, $X'^{m}$ is the generator corresponding to the weight $(-,-,-,+,-,-;0,0)$ of the $\rep{32'}$ representation, with $+$ in the $m$-th position.

In the same way we can obtain the terms on the right hand side. Thus, by acting with  $E_{\alpha_7}$ on the fluxes  we see that
\begin{eqnarray}
P^{12}_ p & \rightarrow &   \omega'^{12}_ p\\
{\tilde H}^{ 12p} & \rightarrow & P'^{p,12}- \frac{1}{\sqrt{2}}\bP^{12p}\\
 \bQ^{i}   & \rightarrow & \bom'^{i}
\end{eqnarray}
whereas the action on the other terms which appear in the right hand side of eq.(\ref{alg1}) vanishes. Putting all together, and proceeding in a similar way for the other $[\hat X^m,X^n]$ commutators, we get
\begin{eqnarray}
{[X'^{m} , \hat X^{n}]}&=& \omega'^{mn}_{p} \hat X^{ p}+(P'^{p,mn} +
\frac{1}{\sqrt{2}}\bP^{mnp}) \hat X_{p}+2\bom'^m \hat X^{n} +\bom'^{n}
\hat X^{m}\\
&&+Q^{ab}_ p X'^{p}+  \bQ^
m X'^{n}+     2\bQ^
n X'^{m}\nonumber
\end{eqnarray}
Following similar steps we could derive the rest of commutators.
For instance,  by acting with the generator corresponding to the negative root  $\mu  \equiv  (-,-,-,-,-,-;+,-)$ we would have
$ E_{\mu}: {[X'^{1}, \hat X^{2}]} \rightarrow {[X'^{1}, X'^{2}]}$  in the ${\bf 32}'\times {\bf 32}' $.
The presentation of the full algebra and constraints is beyond the scope of this work.

\subsection{Toroidal orbifolds and flux induced superpotentials}

Whereas compactification on a $T^6$ orientifold leads to $\mathcal{N}=4$ theories in 4d, one can easily reduce the amount of supersymmetry to $\mathcal{N}=1$ by orbifolding the theory with a discrete symmetry group $\Gamma \subset SU(3)$. Toroidal orbifolds are the simplest examples of compactifications with global $SU(3)$ structure. The general (non toroidal) case will be considered in Section \ref{sec4}.

The feature that makes toroidal orbifolds quite treatable despite the small amount of supersymmetry preserved, is that the structure group of the tangent bundle is still the trivial one, except at the orbifold singularities, where it is reduced to $SU(3)$. Hence, we can distinguish two types of states in the effective theory: \emph{untwisted states}, which arise from direct truncation of the parent $\mathcal{N}=4$ theory and are invariant under a global symmetry group $G\subset \Tsub$, and \emph{twisted states} localized at orbifold singularities, which are not directly related to a truncation of the parent $\mathcal{N}=4$ theory and transform under a larger symmetry group $G_{twist} \not\subset E_7$.

In this way, we can easily make use of the global $\Tsub$ symmetry of the parent $\mathcal{N}=4$ theory to describe the effective action of untwisted fields, keeping in mind that only the subset of fields invariant under $\Gamma$ survive in the orbifolded $\mathcal{N}=1$ theory.

\subsubsection{Type IIB orientifold compactifications}

We focus here on type IIB compactifications on $T^6/[\Omega_P(-1)^{F_L}\sigma\times \Gamma]$, where the orientifold involution $\sigma$ reverses all coordinates of $T^6$. In particular, $\sigma(J)=J$ and $\sigma(\Omega)=-\Omega$. There are O3-planes spanning the space-time directions and, depending on the particular choice of $\Gamma$, there can be also O7-planes wrapping complex 4-cycles within the $T^6$. Consistently with this, D3 and/or D7-branes may be also required in the compactification to cancel the total RR charge.

Following \cite{kst} we can introduce a basis for 3-forms in the covering $T^6$ as,
\begin{align}
\alpha_0 &= dx^1\wedge dx^2\wedge dx^3 \ , &
\alpha_i{}^{\tj}&=\frac12 \epsilon_{ilm}dx^l\wedge dx^m\wedge dx^{\tj}\ , \\
\beta^0 &= dx^{\tilde 1}\wedge dx^{\tilde 2}\wedge dx^{\tilde 3}\ , & \beta^i{}_{\tj}&=-\frac12 \epsilon_{\tj\tilde l\tilde m}dx^{\tilde l}\wedge dx^{\tilde m}\wedge dx^{i}\ , \qquad i,\tj=1,2,3\nonumber
\end{align}
with
\begin{equation}
\int \alpha_i{}^{\tj} \wedge \beta_{\tilde l}{}^k=\delta_i{}^k \delta_{\tilde l}{}^{\tj}
\end{equation}
In terms of these, the holomorphic 3-form can be expanded as,
\begin{equation}
\Omega=\alpha_0+U^i{}_{\tj}\alpha_i{}^{\tj}-(\textrm{cof }U)_i{}^{\tj}\beta^i{}_{\tj}+(\textrm{det }U)\beta^0
\label{omega}
\end{equation}
where,
\begin{equation}
(\textrm{cof }U)_i{}^{\tj}=\frac12\epsilon_{ilm}\epsilon^{\tj\tilde p\tilde q}U^l{}_{\tilde p}U^m{}_{\tilde q}\ , \qquad \textrm{det }U=\frac{1}{3!} \epsilon_{ilm}\epsilon^{\tj\tilde p\tilde q}U^l{}_{\tilde p}U^m{}_{\tilde q}U^i{}_{\tj}
\end{equation}
The scalars $U^m{}_{\tilde q}$ correspond to moduli of the complex structure defined by $\Omega$ in $T^6$,
\begin{equation}
\Omega=dz^1\wedge dz^2\wedge dz^3\ , \qquad dz^m = dx^m + U^m{}_{\tilde p}dx^{\tilde p}
\end{equation}
Similarly, the moduli parameterizing deformations of the K\"ahler structure of $T^6$ can be extracted from the complexified 4-form,\footnote{We have taken $C_4\to -C_4$ and $\mathcal{J}_c\to -\mathcal{J}_c$ with respect to \cite{acfi,acr} in order to match the usual conventions in Generalized Complex Geometry.}
\begin{equation} \label{Jc}
\mathcal{J}_c=C_4-\frac{i}{2}e^{-\phi} J\wedge J=-T^{l\tilde m}\omega_{l\tilde m}\ ,
\end{equation}
where $\omega_{l\tilde m}$ is a basis of integer 4-forms even under the orientifold
involution,\footnote{\label{notatoro}Notice that before applying the orbifold
projection $\Gamma$, the first of the equations in (\ref{JOmega}) below is
generically not satisfied, and it is only after the projection that $J$ and
$\Omega$ define an SU(3) structure.}
\begin{equation}
\omega_{i\tj}=\frac14\epsilon_{ilm}\epsilon_{\tj \tilde p \tilde q}dx^l\wedge  dx^{\tilde p}\wedge dx^m\wedge dx^{\tilde q}
\end{equation}
The complex axion-dilaton $S_B$ has been defined in eq.(\ref{iibS}).
The complex scalars $U^m{}_{\tilde q}$, $T^{l\tilde m}$ and $S_B$ parameterize the coset $\frac{SL(2)|_B}{U(1)} \times \frac{O(6,6)}{O(6)\times O(6)}$. Their weight vectors can be obtained from the adjoint representation of $E_7$, summarized Table \ref{ta:133}. For that one has to note that only the $(\mathbf{66},\mathbf{1})+(\mathbf{1},\mathbf{3})\subset\mathbf{133}$ of $\Tsub$ survive the orientifold projection, according to what was described in Section \ref{sec:e7break}. Subtracting out the $O(6)\times O(6)$ and $U(1)$ pieces associated to local gauge transformations and keeping elements in the Borel subalgebra, then leads to the  weight vectors in Table \ref{escalar}.

\begin{table}[!ht]
\begin{center}
\begin{tabular}{|c|c|}
\hline
scalar & weight\\
\hline \hline
$S$&$(0,0,0,0,0,0\ ;-1,1)$\\
\hline
$U^m{}_{\tilde q}$&$( \underline{1,0,0},\underline{-1,0,0} \ ;0,0)$\\
\hline
$T^{l\tilde m}$&$(\underline{1,0,0},\underline{1,0,0} \ ;0,0)$\\
\hline
\end{tabular}
\caption{\small{Weights in the $(\mathbf{66},\mathbf{1})+(\mathbf{1},\mathbf{3})$ representation of $O(6,6)\times SL(2,\mathbb{R})|_B$ associated to the moduli of $T^6$.}\label{escalar}}
\end{center}
\end{table}

The metric of this moduli space is given in terms of the K\"ahler potential \cite{gl1},
\begin{equation}
\hat K_{O3}=-\textrm{log}\left[-i (S-\bar S) \right]-\textrm{log}\left[-i\int\Omega\wedge \bar \Omega\right]-2\textrm{log}\left[\frac16\int J\wedge J\wedge J\right]\label{kahler}
\end{equation}
An explicit expression of the second integral in terms of the complex structure moduli for a general $T^6$ can be found, for instance, in \cite{kst}.

We can consider now the effect of switching on background fluxes in the compact manifold.
Let us start turning on standard supergravity fluxes. The ones that survive the orientifold and orbifold projections are RR and NSNS 3-form fluxes along three cycles of the internal $T^6$. The deformation induced in the 4d effective supergravity theory is particularly well-known in this case.  Their effect is encoded in a non-trivial effective superpotential of the form \cite{gvw},
\begin{equation}
W_{O3}=\int\Omega\wedge (F_3-S_BH_3)
\label{gkp}
\end{equation}
Note that, since the covering $T^6$ is parallelizable, there is a basis of globally defined one-forms that can be used to define global fluxes that are non-trivial in cohomology, such that the integral in eq.(\ref{gkp}) does not vanish. Locally, these fluxes are introduced by local gauge fields as discussed in previous sections.

Three-form fluxes generically induce an overall charge of D3-brane. Under global monodromies of the $T^6$, both $C_2$ and $B_2$ shift, and one has to patch the background by using gauge transformations with parameters $\Lambda$ for $B_2$, as in eq.(\ref{Bpatches}), and similarly $\hat\Lambda_1$ for $C_2$.

The superpotential (\ref{gkp}) is at the core of many of the recent phenomenological approaches to string theory, where moduli stabilization is a must. However, in Section \ref{sec2b} it was recast that $F_3$ and $H_3$ are only a piece of the $(\mathbf{220},\mathbf{2})$ representation of $O(6,6)\times SL(2,\mathbb{R})|_B$. In Table \ref{220S} we summarized all the elements in the representation.

We can therefore generalize eq.(\ref{gkp}) to incorporate all the remaining fluxes in the $(\mathbf{220},\mathbf{2})$ representation. Following the same strategy than in \cite{stw,acfi,acr}, we act with the $O(6,6)\times SL(2,\mathbb{R})|_B$ generators on (\ref{gkp}), or rather, on the invariant superpotential, $\mathcal{G}=\textrm{log}|W_{O3}|^2+\hat K_{O3}$, with $\hat K_{O3}$ given in (\ref{kahler}). For that we observe that $W_{O3}$ transforms with weight $( 1,1,1,0,0,0;-\frac12,\frac12)$ and moduli transform with weights summarized in Table \ref{escalar}. We can build $O(6,6)/[O(6)\times O(6)]$ covariant 3-forms linear in the fluxes by considering contractions of the fluxes in Table \ref{220S} with one or more copies of the 4-form $\mathcal{J}_c$. The number of copies of $\mathcal{J}_c$ which are required is dictated by the tensorial structure of the fluxes, derived in previous sections. Moreover, since the holomorphic 3-form $\Omega$ carries no charge of $SL(2)/U(1)|_B$, the combination of fluxes which couples to $\mathcal{J}_c$ has to be always of the form $f_2-Sf_1$, with $f_2$ ($f_1$) the highest (lowest) component in the doublet of $SL(2)|_B$. Based on the above observations, it is possible to show then that the superpotential involving the full $(\mathbf{220},\mathbf{2})$ representation of fluxes is given by,
\begin{equation}
W_{O3}=\int \Omega\wedge [(F_3-S_BH_3)+(Q-S_BP)\cdot\mathcal{J}_c+(P'-S_BQ')\cdot\mathcal{J}_c^2+(H'-S_BF')\cdot\mathcal{J}_c^3]
\label{supgen}
\end{equation}
where,
\begin{align}
(Q\cdot\mathcal{J}_c)_{p_1p_2p_3}&=\frac12Q^{mn}_{[p_1}(\mathcal{J}_c)_{p_2p_3]mn}\nonumber\\
(P'\cdot\mathcal{J}_c^2)_{p_1p_2p_3}&=\frac{1}{4^2\cdot 4!}P'^m_{[p_1 p_2|} ({\cal J}_c)_{i_1 i_2 i_3 i_4} ({\cal J}_c)_{i_5 i_6 m |p_3]} \epsilon^{i_1 ... i_6} \ ,\\
(H'\cdot\mathcal{J}_c^3)_{p_1 p_2 p_3}&=\frac{5}{3!\cdot 128}H'^{p_4p_5p_6}({\cal J}_c)_{i_1 i_2 i_3 i_4} ({\cal J}_c)_{i_5 i_6 [p_5 p_6} ({\cal J}_c)_{p_1 p_2 p_3 p_4]} \epsilon^{i_1 ... i_6}\nonumber
\end{align}
and similarly for $P\cdot\mathcal{J}_c$, $Q'\cdot\mathcal{J}_c^2$ and $F'\cdot\mathcal{J}_c^3$. The reader may check that all terms in this equation indeed transform with weight $( 1,1,1,0,0,0;-\frac12,\frac12)$, as desired.

Of course, one has to bear in mind that in concrete models many of the flux components which appear in (\ref{supgen}) are projected out by the orbifold action $\Gamma$. In particular, one may check that (\ref{supgen}) reproduces the results derived by similar arguments in \cite{acfi,acr} for the case of $\Gamma=\mathbb{Z}_2\times\mathbb{Z}_2$ and $G=SL(2,\mathbb{R})^7$.\footnote{For completeness we present here the embedding of $SL(2,\mathbb{R})^7\subset E_7$ selected by the $\mathbb{Z}_2\times\mathbb{Z}_2$ orbifold. This is given by 7 coordinate vectors, $\sigma_i$, one per $SL(2,\mathbb{R})$ factor. In the conventions of \cite{acfi} for $SL(2,\mathbb{R})^7$, these are given by,
\begin{equation*}
\sigma_0=\frac12(\tilde e_8-\tilde e_7)\ , \quad \sigma_i=\frac12(\tilde e_i+\tilde e_{i+3})\ , \quad \sigma_{i+3}=\frac12(\tilde e_i-\tilde e_{i+3})\ ,\qquad i=1,2,3
\end{equation*}
with $\tilde e_i$ the orthonormal vectors of the $B$ basis, introduced in Section \ref{sec:e7break}.} In addition, the quadratic constraints derived in Section \ref{sec3} require in general further components to vanish. Hence, many of the vacua which result from (\ref{supgen}) are actually related by 4d electric-magnetic duality. Systematic analysis of the the vacuum structure induced by the first two pieces in the above superpotential  for $\Gamma=\mathbb{Z}_2\times\mathbb{Z}_2$ have been carried out recently in \cite{gmd}.

We could consider also fluxes transforming in the $(\rep{12},\rep{2})$ representation of $\Tsubb$, since they survive to the orientifold projection too, by acting with $E_7$ generators on eq.(\ref{supgen}). We postpone however the discussion of these fluxes to Section \ref{secgen}, where the complete flux induced effective superpotential is derived in the broader context of general $SU(3)$ structure compactifications.

\subsubsection{Heterotic orbifold compactifications}

The derivation of the flux induced effective superpotential for heterotic compactifications on toroidal orbifolds follows closely the discussion in the preceding subsection. Complex structure moduli are still defined in terms of the holomorphic 3-form, as in eq.(\ref{omega}). On the other hand, K\"ahler moduli are now given in terms of the complexified 2-form,
\begin{equation}
J_c=B_2+iJ=T_{l\tilde m}\hat\omega^{l\tilde m}
\end{equation}
where $\hat\omega^{l\tilde m}$ is a basis of integer 2-forms (see footnote \ref{notatoro}), $\hat\omega^{l\tilde m}=dx^l\wedge dx^{\tilde m}$. The heterotic axion-dilaton $S_H$ was defined in eq.(\ref{hetS}). These definitions are such that the corresponding weight vectors in the $\Tsubh$ basis are still given by Table \ref{escalar}. All together, the scalars parameterize the coset $\frac{SL(2)|_H}{U(1)} \times \frac{O(6,6)}{O(6)\times O(6)}$ with K\"ahler metric,
\begin{equation}
\label{khet}
\hat K_{\rm het.}=-2\textrm{log}[-i(S_H-\bar S_H)]-\textrm{log}\left[-i\int\Omega\wedge \bar \Omega\right]-\textrm{log}\left[\frac16\int J\wedge J\wedge J\right]
\end{equation}

To get the full superpotential, the starting point is the known effective superpotential for heterotic compactifications with 3-form NSNS flux and torsion \cite{heterotic1,heterotic2,heterotic3,heterotic4}
\begin{equation}
W_{\rm het.}=\int \Omega\wedge (H_3+i dJ)
\end{equation}
We can now proceed as before in order to extend this superpotential to the full set of fluxes transforming in the $(\mathbf{220},\mathbf{2})$ representation of $O(6,6)\times SL(2,\mathbb{R})|_H$. These are summarized in the last column of Table \ref{220S}. After some algebra we arrive to the expression,
\begin{equation}
W_{\rm het.}=\int \Omega\wedge\left[(H_3-S_H \tilde H_3)+(\omega -S_H Q') \cdot J_c+(Q-S_H \omega') \cdot J_c^2+(R-S_H H') \cdot J_c^3\right] \label{suphet}
\end{equation}
where the contractions are defined as,
\begin{align} \label{contrac}
(\omega \cdot J_c)_{p_1p_2p_3}&=\omega^m_{[p_1p_2}(J_c)_{p_3]m} \nn \\
(Q\cdot J_c^2)_{p_1p_2p_3} &=\frac12 Q^{mn}_{[p_1}(J_c\wedge J_c)_{p_2p_3]mn}\\
(R\cdot J_c^3)_{p_1p_2p_3} &=\frac{1}{3!} R^{p_4p_5p_6}(J_c\wedge J_c\wedge J_c)_{p_1p_2p_3p_4p_5p_6} \nn
\end{align}
and similarly for $Q'$, $\omega'$ and $H'$ (c.f. footnote \ref{footnot}).

It is illuminating to express this superpotential in terms of the tensors $f^{ABCi}$ and $f^{Ai}$, defined in eqs.(\ref{fluxes332220}), which in the present context are just the embedding tensors of the parent $\mathcal{N}=4$ gauged supergravity theory \cite{sw}. We find the compact expression
\begin{equation}
W_{\rm het.}= \int \Omega\wedge [{\bbG}_3  \cdot e^{J_c}+{\bbG}_1  \cdot \ e^{J_c}] \ .
\end{equation}
where we have included also fluxes transforming in the $(\rep{12},\rep{2})$ representation whose superpotential will be derived in Section \ref{secgen}.
${\bbG}_3$ and ${\bbG}_1$ are defined as
\begin{equation}
 ({\bbG}_3)_{ABC}= f_{ABC2}-S_Hf_{ABC1}\ , \qquad  ({\bbG}_1)_{A}= f_{A2}-S_Hf_{A1}
\end{equation}
where a subindex $2$ ($1$) refers to the direction $v_i$ ($\omega_i$),
and the contraction $\cdot$ is the $O(6,6)$ action $f_{ABCi} \Gamma^{ABC}$ and $f_{Ai} \Gamma^{A}$, with the gamma matrices acting as in eq.(\ref{Gamma}).

\section{General $\mathcal{N}=1$ compactifications}
\label{sec4}

In this Section we consider compactifications that preserve $\mathcal{N}=1$ supersymmetry in 4d, and which are not necessarily toroidal. From the phenomenological point of view this is perhaps the most appealing case. The existence of a single nowhere vanishing spinor reduces the structure  of the tangent bundle to a global $SU(3)$ structure. The latter can be completely characterized in terms of a globally defined $SU(3)$ invariant $(1,1)$-form $J$, and a holomorphic $(3,0)$-form $\Omega$, which satisfy the relations
\begin{equation} \label{JOmega}
J\wedge \Omega = 0\ , \qquad J\wedge J\wedge J = -i\frac34 \Omega\wedge\bar \Omega
\end{equation}
In what follows we first review the $O(6,6)$ and $E_7$ covariant formulations of $\mathcal{N}=1$ backgrounds, following respectively \cite{GLW,BG} and \cite{E7}. Then we address in Section \ref{secgen} the computation of the effective superpotential in general $SU(3)$ structure compactifications with arbitrary fluxes.

\subsection{$O(6,6)$ formulation of $\mathcal{N}=1$ backgrounds}

Four-dimensional $\mathcal{N}=1$ supersymmetry requires the existence of nowhere vanishing internal spinors $\eta^1$ and $\eta^2$ such that
the 4d supersymmetry parameter $\varepsilon$ is obtained from the two 10d ones $\epsilon^{1,2}$
by
\begin{equation}
\begin{aligned}
\label{decompepsilon}
   \epsilon^1 &= \varepsilon_+ \otimes \eta^1_-
      + \varepsilon_- \otimes \eta^1_+ \ , \\
   \epsilon^2 &= \varepsilon_+ \otimes \eta^2_-
      + \varepsilon_- \otimes \eta^2_+ \ ,
\end{aligned}
\end{equation}
We have chosen the chirality of the 10d spinors to be negative. There is no requirement a priori on the relative
orientation of the spinors $\eta^{1,2}$.
Each of them is invariant under
an $SU(3)$ structure on the 6-dimensional space. If the two spinors are the same, they give rise to a single $SU(3)$ structure.
Whenever the spinors do not coincide, the two $SU(3)$ structures intersect into an $SU(2)$. If there are points on the manifold where the spinors are parallel, there is no global $SU(2)$ structure, but only a local one.

The two spinors can be combined
to form complex pure spinors of $O(6,6)$
\begin{equation}
\label{purespinors}
   \Phi_0^+ =  \eta^1_+ {\eta}^{2\, \dagger}_+
     \ , \qquad
   \Phi_0^- =   \eta^1_+ {\eta}^{2\, \dagger}_-  \  .
\end{equation}
Making use of Fierz identities, it is possible to express $\Phi_0^{\pm}$ as sums of spinor bilinears, namely
\begin{equation}
\label{eq:bispinors}
   \eta^1_+ \eta^{2 \, \dagger}_+ = \frac{1}{8}\sum_{p}
      \frac{1}{p!}\left(
         {\eta}^{2\, \dagger}_\pm \gamma_{m_1\dots m_p}\eta^1_+\right)
         \gamma^{m_p\dots m_1} \, .
\end{equation}
By chirality, only even (odd) $p$ contribute to $\Phi^+$ ($\Phi^-$) and these can equivalently be thought of as sums of even or odd forms in the $\rep{32}$ and $\rep{32'}$ representations of $O(6,6)$.
In the special case
where the two spinors coincide, i.e. $\eta^1=\eta^2$, the pure spinors read
\beq \label{JOmegaphi}
\Phi_0^+=e^{-iJ} \ , \qquad \Phi_0^-=-i \Omega \ ,
\eeq
where $J$ and $\Omega$ are those of eqs.(\ref{JOmega}).

By construction $\Phi^\pm_0$ are pure,  i.e. each of them is  annihilated by half of the twelve  $O(6,6)$
gamma  matrices $\Gamma^{A}$, splitting the generalized tangent bundle into  a 6-dimensional  complex holomorphic bundle (given by $i_\pm=1,...,6$ such that  $\Gamma^{i_\pm} \Phi^\pm_0=0$) and its complex conjugate antiholomorphic bundle.\footnote{Furthermore, $\Phi^\pm_0$ are by construction
compatible, which implies that the two 6-dimensional holomorphic bundles have a 3-dimensional intersection.}
In other words, $\Phi^\pm_0$ define each
a generalized (almost) complex structure ${\cal J}_0^\pm$, satisfying $({\cal J}_0^\pm)^2=-\id$, that can be obtained from the spinors
by
\begin{equation}
\label{Jgen}
   \mathcal{J}_0^{\pm A}{}_B\ = i \, \frac{
       \mukai{\Phi^\pm_0}{\Gamma^A{}_B \bar{\Phi}^\pm_0}}
       {\mukai{\Phi^\pm_0}{\bar{\Phi}^\pm_0}}\ ,
\end{equation}
where $O(6,6)$ gamma matrices act as in (\ref{Gamma}), and the bracket is the Mukai pairing defined in eq.(\ref{mukai}), which
is the natural bilinear on $O(6,6)$ spinors.

Two compatible pure spinors define a generalized (positive definite) metric on the generalized tangent space.
Using the generalized almost complex structures associated to the pure spinors, the generalized metric is given by
\beq
{\cal H}=-\eta \, \cj^+ \cj^- \ .
\eeq
This is the same metric as in eq.(\ref{eq:genmetric}). For the pure spinors (\ref{JOmegaphi}), it gives a block diagonal
matrix (corresponding to $B_2=0$), where $g$ is obtained in the standard way from an $SU(3)$ structure (in complex coordinates $g_{i \bj}=-i J_{i \bj}$).

In order to obtain a generalized metric that contains a $B_2$ field, one needs to
$B$-transform the pure spinors, i.e. to apply an $O(6,6)$ transformation corresponding to a $B_2$ field to eq.(\ref{JOmegaphi})
(or more generally to eq.(\ref{purespinors})). Making use of eq.(\ref{adjac}) and the representation for gamma matrices in eq.(\ref{Gamma}),
 it is not hard to see that the $B$-transform on the
pure spinors amounts to a wedge product of $B_2$ and the component forms. Exponentiating the action we get $ \Phi^\pm=e^{B_2} \Phi_0^\pm$ which for the case  $\eta^1=\eta^2$, corresponding to a single SU(3) structure, implies
\beq
 \Phi^+=e^{B_2-iJ} \ , \qquad \Phi^-=-i e^{B_2} \Omega \ .
 \eeq
In terms of the pure spinors, the K\"ahler potential for the $\mathcal{N}=1$ theory in the context of type IIB compactifications reads \cite{BG}
\beq
\hat K=-\log \left[ i \int \mukai{\Phi^-}{\bar \Phi^-} \right] - 2\log  \left[i\int \mukai{e^{-\phi} \Phi^+}{e^{-\phi} \bar \Phi^+} \right] \ .
\label{genkahler}
\eeq
It is easy to see that this is equivalent to (\ref{kahler}). Note that the contribution of $B_2$ drops out, as it should be since the Mukai pairing is an $O(6,6)$ invariant, and the $B_2$ field is an adjoint $O(6,6)$ action.

The second term in the K\"ahler potential should be written in terms of the $\mathcal{N}=1$ variables ${\cal J}_C$ and $S_B$. These can be read off from a combination of $\Phi^+$, the dilaton and the RR potentials into the following complex form \cite{BG}
\beq \label{PhiC}
\Phi^+_C=e^{B_2} C^+ + i e^{-\phi} \re (e^{i \theta} \Phi^+)
\eeq
where $\theta$ is an angle that defines the $\mathcal{N}=1$ orientifold projection. For O3/O7 planes $\theta=0$, while for O5/O9, $\theta=\pi/2$.
In the case of type IIB compactifications with O3-planes, $\Phi^+_C=e^{B_2} (S_B+{\cal J}_C+C_2+C_6)$. Hence the 0-form component of $\Phi^+_C$ is the complex axion-dilaton $S_B$, while the 4-form component is precisely ${\cal J}_C$ defined
in eq.(\ref{Jc}). Note that even if $B_2$, $C_2$ and $C_6$ are projected out of the spectrum in this case, we keep them in $\Phi^+_C$ as their fluxes are not.

In terms of $\Phi^+_C$ the $O(6,6)$-covariant superpotential is given by
\beq \label{supGCG}
W=  \int \mukai{d  \Phi^+_C}{\Phi^-} \ .
\eeq
Here the fluxes $F_3$ and $H_3$ are encoded in their local definition in terms of the gauge fields, i.e. $F_3=dC_2$, $H_3=dB_2$. It is not hard to check that this reproduces eq.(\ref{gkp}).

The term in eq.(\ref{supgen}) involving the locally geometric NSNS flux $Q^{ij}_k$ can also be easily encoded in (\ref{supGCG}). For that, note that in general $\Phi^+_C=e^{B_2} (\Phi^+_{C})_0$, where $(\Phi^+_{C})_0$ is built out of $C^+$ and the pure spinor $\Phi^+_0$.  If instead of a $B$-transform one acts by a $\beta$-transformation, the action of $d$ on $\beta_2$ gives the $Q \cdot {\cal J}_C$ term in eq.(\ref{supgen}).

The remaining NSNS fluxes $Q'$ and $H'$ in the $\rep{220}$ representation are locally non-geometric and their contribution to the superpotential cannot be obtained from (\ref{supGCG}), not even promoting $d$ to an element in the $\rep{12}$ of $O(6,6)$. The reason for this is that, as explained in Section \ref{sec2b}, these fluxes require
extending the derivatives even further,
going beyond the fundamental representations of $O(6,6)$ to the $\rep{56}$ representation of $\E7$.  Similarly, non-geometric RR fluxes should also only appear when  promoting the $O(6,6)$ covariance to a full $\E7$ one.

\subsection{$E_7$ formulation of $\mathcal{N}=1$ backgrounds}

We review now the $\E7$ covariant formulation of $\mathcal{N}=1$ backgrounds that descend from $\mathcal{N}=2$ ones, following \cite{E7}. We work in the $\Tsubh$ basis, where the $O(6,6)$ subgroup is that of Generalized Geometry.

The first step is to embed $\Phi_0^-$ and $\Phi^+_0$ into representations of $\E7$. The easiest way is to embed them in the $\rep{32'}$ and $\rep{32}$ representations of $O(6,6)$ that appear respectively in the decomposition of the $\rep{56}$ and $\rep{133}$ representations of $\E7$ (c.f. eqs.(\ref{56}) and (\ref{133})).  This assignment is also consistent with the degrees of freedom in the $\mathcal{N}=2$ theory: those in $\Phi^-$
are the scalars of vector multiplets, while the deformations of $\Phi^+$ build up, together with the dilaton and the RR axions, the hypermultiplets.

The moduli space is a direct product of these two moduli spaces, which implies in particular that $\Phi^-_0$ should be a singlet under
$\SLR|_H$.  The pure spinor $\Phi_0^-$ is therefore embedded as follows
\beq \label{lambda0}
\lambda^0=(0,\Phi_0^-) \, \in \rep{56} \ .
\eeq
where we have used the notation in eq.(\ref{56}).

The other spinor, $\Phi_0^+$, combines with the dilaton to form a doublet of $\SLR|_H$. It should also
combine with the RR fields and transform non-trivially under the $SU(2)_R$ R-symmetry. The way to realize all these conditions is to promote $\Phi_0^+$
to an $SU(2)_R$ triplet of elements $(K^0_1, K^0_2, K^0_3)$, with $K^0_a \in \rep{133}$, satisfying the real $su(2)$ algebra $[K^0_a, K^0_b]=2 \kappa \epsilon_{abc} K^0_c$, with $\kappa=e^{-2\phi} {\rm vol}_6 $.  Using the notation in  eq.(\ref{133}), we define $K^0_+=K^0_1+i K^0_2$
by
\beq \label{K0}
K^0_+=\left(0,0,e^{-\phi} (v^i S_H +\omega^i \right)\,  \Phi_0^+) \, \in \rep{133}
\eeq
where $v^i$, $\omega^i$ are those introduced in Section \ref{sec56} and $S_H$ is the heterotic complex dilaton defined in eq.(\ref{hetS}).
$K_-^0$ is just the complex conjugate of $K_+^0$, while $K_3^0$, obtained by demanding the $su(2)$ commutation relations, is given by
\beq \label{K30}
K_3^0=\frac{1}{4}  \left(u^i \bar u_j+\bar u^i u_j, i u \bar u \, {\cal J}_0^{+A}{}_B, 0 \right)
\eeq
with $u\equiv  (v^i S_H +\omega^i) $, $u \bar u=(S-\bar S)_H=2i e^{-2\phi} \textrm{vol}_6$ and ${\cal J}_0^{+A}{}_B$ is the
generalized almost complex structure defined by $\Phi_0^+$ in eq.(\ref{Jgen}).

In an analogous way as for the $b$-transform of $\Phi^\pm$, we can define $b_2$, $b_6$ and $c^+$ transformed objects of $\lambda^0$ and $K^0$ as
\beq
\lambda=e^{c^+} e^{b_6} e^{b_2} \lambda^0 \ , \qquad  K_a=e^{c^+} e^{b_6} e^{b_2} K_a^0
\label{Klambda}
\eeq
where $e^{c^+}$, $e^{b_6}$ and $e^{b_2}$ stand for the adjoint $\E7$ action, given in eq.(\ref{adjac}), by the corresponding generators in (\ref{mugeomH}).
The $b_6$ action on $K_+$ shifts $S_H \to S_H + b_6$, as expected. The $c^+$ action generates, among other terms, a non-zero
last component in $K_3$, proportional to $c^+$. This implies that $(e^{-\phi} \, \re \Phi^+,e^{-\phi}\,  \im \Phi^+,C^+)$ forms a triplet of $SU(2)_R$, as it should.

An $\mathcal{N}=1$ supersymmetry is selected by choosing a $U(1)_R \subset SU(2)_R$, or equivalently a vector $r^a$ such that
a triplet of $SU(2)_R$ decomposes as a singlet and a doublet. The $\mathcal{N}=1$ supersymmetry selected in heterotic compactifications can be parameterized as $r^1=r^2=0$, $r^3=1$, while for type II orientifold compactifications $r^3=0$ and the $\mathcal{N}=1$ supersymmetry is parameterized in this case by a single angle $\theta$ in the $(r^1,r^2)$ plane.
The singlet and doublet components in the triplet $(K_+, K_-, K_3)$ are selected in each case by the vectors
\beq \begin{aligned} \label{z,r}
{\rm heterotic:} \quad (r^+,r^-,r^3)&=  (0,0,1) \ ,  & (z^+,z^-,z^3)&=(1, 0,0) \ , \\
{\rm type \ II:} \quad (r^+,r^-,r^3)&=  (ie^{i\theta},-i e^{-i\theta},0) \ , \  &(z^+,z^-,z^3)&=(\tfrac{i}{2} e^{i \theta}, \tfrac{i}{2} e^{-i\theta},1) \ .
\end{aligned}
\eeq
Hence, defining the $\mathcal{N}=1$ field as
\beq
K_C= z^a K_a \ ,
\eeq
we have that the spinor component along $\omega^i$ in the $(\rep{2,32})$ piece of $K_C$ is indeed the $\mathcal{N}=1$ chiral field in eq.(\ref{PhiC}) for type II compactifications.

\subsection{General U-duality covariant superpotential}
\label{secgen}

We are now ready to compute with the above tools the U-duality covariant superpotential for general $SU(3)$ structure compactifications. In terms of the geometric objects $\lambda$ and $K_C$ the superpotential in the $E_7$ covariant formulation is given by \cite{E7}
\beq \label{WE7}
W=\int {\cal S}(\lambda, D K_C) \ .
\eeq
Here $D$ is the generalized derivative defined in eq.(\ref{deriv56}), transforming in the $\rep{56}$  representation,  $K_C=g K^0_C$ and $\lambda=g \lambda^0 $, with $g=e^{A} $ a generic group element of $E_7$.
 ${\cal S}$ is the symplectic invariant in the $\rep{56}$ representation, whose $\Tsub$ decomposition is given in eq.(\ref{sympl}) of the Appendix and  $z^a$ is the vector introduced in (\ref{z,r}). For consistency $D K_a$, the generalized derivative of $K_a$, is projected onto the $\rep{56}$ representation.

Extracting the group elements $g$ and making use of the symplectic invariance, the superpotential above can be recast as
\beq \label{WE70}
W=\int {\cal S}(\lambda^0, {\cal D} K^0_C)
\eeq
where  the derivative is now acting on the bare objects with a superindex $0$, while   derivatives of the gauge fields (the fluxes) have been encoded in the generalized connection \cite{E7,GO},
\beq \label{gencon}
{\cal D}^{\cal AB}{}_{\cal C} = D^{\cal A} \delta^{\cal B}{}_{\cal C} + F^{\cal AB}{}_{\cal C} \ ,
\eeq
with
\beq
F^{\cal AB}{}_{\cal C}=(g^{-1})^{\cal B}{}_{\cal D} D^{\cal A} g^{\cal D}{}_{\cal C} \ .
\label{gencon2}
\eeq
While in Ref.\cite{E7} the derivative was restricted to the standard derivative, $D=(\partial_m v^i , 0)$, and at the same time the gauge fields were taken to be in the geometric subgroup, here we consider the full $D$ defined in eq.(\ref{deriv56}) and the full set of 133 gauge fields. Hence, the tensor $F^{\cal AB}{}_{\cal C} \in \rep{912}$ involves all the field strengths introduced in Section \ref{sec2b}.

The generalized connection applied to $K^{\cal CD}$ in the $\rep{133}$ and projected onto the $\rep{56}$ gives
\beq \label{covder}
({\cal D}K)^{\cal A}=  {\cal S}_{\cal BC} \left( {D}^{\cal B} K^{\cal CA} + F^{\cal AB}{}_{\cal E} K^{\cal CE} \right) \ .
\eeq
Note that $\lambda^0$, given in eq.(\ref{lambda0}), has only a $(\rep{1,32'})$ piece (equal to $\Omega$ for $SU(3)$ structure compactifications) and therefore only the 3-form piece in the $(\rep{1,32'})$ part of
$({\cal D}K)^{\cal A}$ is kept.
Thus we get
\beq \label{WO3E7}
W=\int \mukai{\Omega}{ {\cal D} K_{C}^{\footnotesize (\rep{1,32'})}} \ .
\eeq

According to our above discussion, for heterotic string compactifications with global $SU(3)$ structure, $K_C=K_+$, defined in eq.(\ref{K0}), and $\Phi_0^+=e^{iJ}$. Fluxes surviving the projection are in the $(\rep{220},\rep{2})$ and $(\rep{12},\rep{2})$ representations of $\Tsubh$. They act on $K^0_+$, which has only a spinor component, as shown in eq.(\ref{912x133}), leading to\footnote{We have canceled the ${\rm vol_6}$ factor in $S_H$ with inverse volume factors in the $\omega^i$ components of the fluxes (see eq.(\ref{fluxes332220})).}
\beq \label{suphetint}
W_{\rm het.}=\int e^{-\phi}\Omega \wedge\left[ (f^{ABC1}   -S_H f^{ABC2})  \Gamma_{ABC} e^{iJ} + (f^{A1}   -S_H f^{A2})  \Gamma_{A} e^{iJ} \right] \ .
\eeq
This is almost exactly the same expression than the one that we derived for toroidal heterotic orbifold compactifications, eq.(\ref{suphet}), except that we are missing the complexification of $J$.  A bare gauge field like $B_2$ appears because the connection $F^{\cal AB}{}_{\cal C}$ in (\ref{gencon2}) is actually defined in terms of generalized derivatives of the
 group elements $g\in \E7$, obtained by exponentiating the generators $A$. In Sections \ref{sec1} and \ref{sec2b} we have actually used only the first order term in the exponentials  to label the fluxes, namely  $H_3$, $\omega$, $Q$, $R$, etc. were defined as
$f^{\cal AB}{}_{\cal C}= D^{\cal A} A^{\cal B}{}_{\cal C}$. The difference between these two definitions involves terms containing bare gauge fields. For the simpler case of heterotic compactifications, where the RR fields are set
to zero, these fluxes differ precisely by factors of $e^{B_2}$, which combine with $e^{iJ}$ to form
$e^{J_c}$.

The other difference with respect to eq.(\ref{suphet}) is an extra overall $e^{-\phi}$ factor in eq.(\ref{suphetint}). This factor can be actually absorbed by a K\"ahler transformation of (\ref{genkahler}), leading to the canonical K\"ahler metric for heterotic string compactifications, given in eq.(\ref{khet}).

Hence, taking all these observations into account, we have that the flux induced superpotential, derived from eq.(\ref{WE7}), for general heterotic compactifications with global $SU(3)$ structure is given by,
\begin{multline}
W_{\rm het.}=\int \Omega\wedge\left[(H_3-S_H \tilde H_3)+(\omega -S_H Q') \cdot J_c+(Q-S_H \omega') \cdot J_c^2+(R-S_H H') \cdot J_c^3\right]\\
+\int \Omega\wedge\left[(\bom-S_H \bQ')\wedge J_c+(\bQ -S_H \bom') \cdot J_c^2\right]\label{whetfin}
\end{multline}
where contractions were defined in eqs.(\ref{contrac}),
\begin{equation}
(\bQ \cdot J_c^2)_{pqr}=Q^m(J_c\wedge J_c)_{pqrm}\label{unac}
\end{equation}
and similarly for $\bom'\cdot J_c^2$.

In a similar way we can compute the superpotential for type IIB compactifications with $O3$-planes. This requires a little more work. The orientifold projection sets in this case
\beq
\theta=0 \ , \quad \Rightarrow \quad z^a K_a=\tfrac{i}{2} (K_++K_-) +K_3 \equiv K_{\rm{O}3} \ .
\eeq
We need to express $\lambda^0$ and $K^0$, given in eqs.(\ref{lambda0}), (\ref{K0}) and (\ref{K30}) in the $\Tsubh$ basis, in terms of the $\Tsubb$ basis.
This can be done with the help of Tables \ref{ta:133} and \ref{ta:56}. The resulting expressions in the $\Tsubb$ basis are
\beq
\begin{aligned}
\lambda^0|_B&=(0, \Omega) \ , \\
K_{\rm{O}3}|_B &={\rm vol_6} \left(-\frac12 e^{-2\phi}  ({\cal J}^+)^m{}_n \ , \ 0 \ , \ \tilde v^i \left(2 e^{-2\phi} ({\cal J^+})^{mn} + e^{-\phi} \epsilon^{123456} -i e^{-3\phi} J + e^{-4\phi} {\rm vol_6} \right) \right. \\
& \qquad \qquad  \left. + \, \tilde \omega^i \left(-\frac{i}{2} e^{-\phi} J^2 \epsilon^{123456} + \epsilon^{123456} -2 e^{-2\phi} ({\cal J}^+)_{mn} + \frac{i}{6} e^{-3\phi} J^3\right) \right)
\end{aligned}
\eeq
where ${\cal J}^+$ is defined in eq.(\ref{Jgen}).

The fluxes surviving the projection are in the $(\rep{2},\rep{12})$ and $(\rep{2},\rep{220})$ representations of $\Tsubb$ (see Table \ref{12S}). In the notation of eq.(\ref{912}), they read
\beq
\begin{aligned}
f^{iA} &= \tilde v^i (\bQ^a + \bP_a \epsilon^{123456}) + \tilde \omega^i (\bP^a + \bQ'_a \epsilon^{123456})   \ .\\
f^{iABC}  &= \tilde v^i \left(F_{abc} +Q^{ab}_c+ P'^a_{bc} \epsilon^{123456} + H'^{abc} \epsilon^{123456} \right) \\
&\qquad \qquad + \tilde \omega^i \left( H_{abc}+P_{a}^{bc}  +Q'^a_{bc} \epsilon^{123456} +F'^{abc} \epsilon^{123456} \right)  \ .
\end{aligned}
\eeq
Making use of ${\rm vol_6}=J^3/6$ and the fact that $({\cal J}^+)_{mn}$ and $({\cal J}^+)^{mn}$ are respectively proportional to $J_{mn}$ and $J^{mn}$,
we obtain
\beq
\begin{aligned} \label{Wmed}
W_{O3}&= \int \Omega \wedge \left[ (F_3-ie^{-\phi} H_3)+(Q-ie^{-\phi}P) \cdot (-\frac{i}{2} e^{-\phi} J^2) \right. \\
& \quad  +(P'-i e^{-\phi} Q') \cdot e^{-2\phi} J
- (H'-i e^{-\phi} F') \cdot (-\frac{i}{6}e^{-3 \phi} J^3) \\
&\qquad \left. +(\bQ^m-ie^{-\phi}\bP^m) \cdot (-\frac{i}{2} e^{-\phi} J^2) + (\bP_m-i e^{-\phi} \bQ'_m) \wedge e^{-2\phi} J\right] \ ,
\end{aligned}
\eeq
where the contractions are as in eqs.(\ref{contrac}) for the $(\rep{220},\rep{2})$ fluxes, while ${\bQ}^m$ and $\bP^a$ act by a single contraction, as in (\ref{unac}).

We proceed now to express (\ref{Wmed}) in terms of the $\mathcal{N}=1$ variables $S_B$ and ${\cal J}_c$ defined in (\ref{iibS}) and (\ref{Jc}) respectively. For that aim, we observe that $J$ and $J^3$, which build up $\textrm{Im}(\Phi^+_0)$, are given in terms of
 $\textrm{Re}(\Phi^+_0)$ by the derivative of the Hitchin functional \cite{hg,GLW}. This leads to the useful relations
\beq
\begin{aligned} \label{JtoJc}
J_{p_1p_2} =\frac{1}{384} J^2_{i_1 i_2 i_3 i_4} J^2_{i_5 i_6 p_1p_2} \epsilon^{i_1 ... i_6} \ , \\
J^3_{p_1 ... p_6} =\frac{5}{128} J^2_{i_1 i_2 i_3 i_4} J^2_{i_5 i_6 [p_1 p_2} J^2_{p_3 p_4 p_5 p_6]} \epsilon^{i_1 ... i_6}
\end{aligned}
\eeq

Note that eq.(\ref{Wmed}) is also missing the contribution from the RR axions
$C_0$ and $C_4$. The reason is the same than for the heterotic superpotential. For instance,
notice that when considering higher orders in the exponential $e^A$, eq.(\ref{adjac}) tells us that the group element corresponding to the 2-form part along $\tilde v^i$ (whose generator is $c_2$) gets a shift $c_2 \to c_2 + c_0 b_2$. Hence,  $d(c_2-c_0 b_2)=F_3 -c_0 H_3$ and the factor of $c_0 H_3$ combines with $i e^{-\phi} H_3$
in eq.(\ref{Wmed}) to build up $S_B H_3$.  When considering only geometric gauge fields, this is the only contribution of the RR axions appear. However, when all fields in the $\rep{133}$ are taken into account, there are extra terms where $C_4$ appears linearly, quadratically or cubically, which combine with $J^2$ factors to build up different powers of ${\cal J}_c$.

Taking into account these two facts, it is not hard to show that  the flux induced superpotential derived from eq.(\ref{WE7}) for general type IIB orientifold compactifications with $O3$-planes and global $SU(3)$ structure is,
\begin{multline}
W_{O3}=\int \Omega\wedge [(F_3-S_BH_3)+(Q-S_BP)\cdot\mathcal{J}_c+(P'-S_BQ')\cdot\mathcal{J}_c^2+(H'-S_BF')\cdot\mathcal{J}_c^3]\\
+\int \Omega\wedge\left[ (\bP_m-S_B \bQ'_m) \wedge \mathcal{J}_c + (\bQ^m-S_B\bP^m) \cdot \mathcal{J}_c^2 \right]\label{wo3fin}
\end{multline}

Let us stress that expressions (\ref{whetfin}) and (\ref{wo3fin}) have been derived in the context of general $SU(3)$ structure compactifications, and therefore not only apply to toroidal orbifold compactifications, but also to general $\mathcal{N}=1$ compactifications of string theory.

It is also interesting to comment on the possible uplift of superpotential (\ref{wo3fin}) to F-theory. Indeed, superpotential (\ref{gkp}) has a well-known interpretation in terms of F-theory compactifications on elliptically fibered Calabi-Yau 4-folds \cite{gvw}. From this point of view, NSNS and RR 3-form fluxes correspond to different components of the M-theory/F-theory 4-form,
\begin{equation}
G_4= \sum_{i=1,2}H_i\wedge\theta^i = (F_3-SH_3)\wedge d\bar z + (F_3-\bar SH_3)\wedge dz
\end{equation}
where $\theta^{i}$ is a basis of integral 1-forms in the elliptic fiber, which we have complexified in the r.h.s. of this expression. In terms of $G_4$, the superpotential (\ref{gkp}) becomes,
\begin{equation}
W=\int \Omega_4\wedge G_4
\end{equation}
where $\Omega_4$ is the holomorphic 4-form, whose existence and uniqueness is guaranteed by the $SU(4)$ holonomy of the Calabi-Yau 4-fold.

Similarly, the second piece in eq.(\ref{wo3fin}) also admits a natural uplift to F-theory.
This term is related to a background for the field strength of the complex parameter (\ref{complexbeta}).
As it has been noted in \cite{acr}, topologically non-trivial $\beta_2$ and $\gamma_2$-deformations (that is, $Q^{ij}_k$ and $P^{ij}_k$ fluxes) correspond to locally geometric type IIB backgrounds with a holomorphic complex dilaton on which there is a deficit of $(p,q)$ 7-brane charge. Thus, generic transition functions gluing different patches contain not only diffeomorphisms and shifts of the $B$ and $C$ fields, but also T-dualities and/or $SL(2)|_B$ rotations.

The form of (\ref{wo3fin}) suggest that $\beta_2$ and $\gamma_2$ are different components of a 1-form along the F-theory fiber,
\begin{equation}
\mathcal{B}^{pq}=\sum_{i=1,2}\mathcal{F}^{pq}_i \theta^i = (\beta^{pq}-S\gamma^{pq}) d\bar z + (\beta^{pq}-\bar S\gamma^{pq}) d z
\end{equation}
so that the first two terms in eq.(\ref{supgen}) are recast as,
\begin{equation}
W=\int\Omega_4\wedge [G_4+d\mathcal{B}\cdot \mathcal{J}_c]
\end{equation}
Note also that taking $\partial_{U^m{}_{\tilde q}}W = W =0$ (in absence of 3-form fluxes), for all complex structure moduli, requires $(Q-S_BP)\cdot\mathcal{J}_c$ to be a primitive $(2,1)$-form. This fact, together with the condition obtained from imposing $\partial_{T^{m\tilde q}}W=0$ for all K\"ahler moduli, implies that (\ref{complexbeta}) has to be an anti-holomorphic $(0,2)$ bi-vector in order to satisfy the $\mathcal{N}=1$ supersymmetry equations.

\section{Conclusions}
\label{sec5}

We have performed a detailed analysis of the relation between Exceptional Generalized Geometry and 4d gauged supergravities, providing an organizing principle for generic string flux compactifications. The U-duality group, $E_7$, which turns out to be also the structure group of the generalized bundle in EGG, encodes much information on the 10d origin of 4d gaugings. In particular, we have established a precise dictionary between weights of the $\mathbf{912}$, $\mathbf{133}$ and $\mathbf{56}$ representations of $E_7$ on one side, and fluxes, gauge fields and gauge parameters, respectively, on the other. Moreover, from different ways of decomposing $E_7$ into $O(6,6)\times SL(2,\mathbb{R})$, we have identified different orientifold projections in 10d, generalizing the results of \cite{ferrara}.

The Clebsh-Gordan decomposition of the $\mathbf{912}$ representation onto the tensor product $\mathbf{56}\times \mathbf{133}$ has provided us also with explicit local expressions for the field strengths in terms of the gauge potentials. This in particular allows to disentangle fluxes which admit an uplift to 10d supergravity backgrounds, from locally non-geometric fluxes, whose field strengths involve exotic exterior derivatives and which should be understood as 4d gaugings resulting from String Theory backgrounds which do not admit a higher dimensional supergravity limit. Hence, the uplift of these backgrounds to 10d should be thought directly in terms of String Theory compactifications. In addition, the $E_7$ group structure allows us also to distinguish globally geometric from globally non-geometric compactifications.

In this way, we have identified systematically all gaugings of 4d supergravity which admit an uplift to backgrounds of 10d supergravity. Some of these backgrounds were already known, corresponding to fluxes of the NSNS and RR field-strengths, metric fluxes or  $\beta$-deformed backgrounds. Our analysis, however, reveals other types of 10d supergravity backgrounds apart from the above ones. These turn out to be related to the RR counterparts of the $\beta$-deformation, which we have dubbed $\gamma$-deformations. We have formulated $\gamma$-deformations in a precise way in the context of EGG. These new deformations provide the natural generalization of the $\beta$-deformation to full-fledged F-theory backgrounds. These results may have also interesting implications, via the AdS/CFT correspondence, for the study of marginal deformations of $\mathcal{N}=4$ Super Yang-Mills with complex $\beta$ parameter \cite{ls}.

The above analysis gives also some clues on the possible 10d origin of some of the particularly less understood gaugings of 4d supergravity. In particular, we have shown that 4d gaugings of $\mathcal{N}=4$ supergravity transforming in the $(\mathbf{12},\mathbf{2})$ representation of
$O(6,6)\times SL(2,\mathbb{Z})$ can be uplifted to backgrounds of 10d type IIB supergravity, corresponding to $\beta$ and/or $\gamma$-deformed backgrounds. Our results are also consistent with \cite{marios}, where it was noticed that a small subset of these gaugings can be also interpreted as particular Scherk-Schwarz reductions (metric fluxes) of heterotic supergravity.

Whereas the above results directly apply to toroidal orbifold or orientifold compactifications, where in absence of fluxes the bulk preserves $\mathcal{N}=8$ or $\mathcal{N}=4$ supersymmetry, they are also relevant in more generic setups. The reason is that even if in generic $\mathcal{N}=1$ compactifications the global structure group is reduced to $SU(3)$, the structure group of the generalized tangent bundle is still given by $O(6,6)\times SL(2,\mathbb{R})$. Hence, in the last part of this work we have made use of the tools of EGG to derive the effective superpotential induced by the fluxes in general $\mathcal{N}=1$ heterotic compactifications or type IIB compactifications with O3 and/or O7-planes. These superpotentials, which we have derived here in a 10d context, allow for the study of the F-term conditions associated to general backgrounds, in particular those which involve $\beta$ and $\gamma$-deformations apart from standard fluxes. We hope to come back to this point in future work.


\section*{Acknowledgments}
{We are very grateful to M. Petrini for collaboration in the early stages of this project. We also thank S. Capriotti, H. Montani, H. Samtleben, P. Vanhove and D. Waldram for useful discussions and comments. P.G.C. and G.A. thank the Kavli Institute for Theoretical Physics and together with M.G thank the Galileo Galilei Institute for hospitality during the completion of this work. This work was partially supported
by MINCYT (Ministerio de Ciencia,
Tecnolog\'\i a e Innovaci\' on Productiva of Argentina) and  ECOS-Sud France binational collaboration project A08E06.}


\appendix

\section{Summary of group theory results on $\E7$}
\label{appa}

\subsection{$\E7$ root and weight system}
\label{app:roots}

$\E7$ has rank seven (7 Cartan generators) and dimension 133. In the Cartan-Weyl basis, the commutation relations of the algebra read,
\begin{equation}
[H^I,H^J]=0\ , \qquad
[H^I,E_\alpha]=\alpha^IE_\alpha\ , \qquad
[E_\alpha,E_\beta]=\begin{cases}\epsilon_{\alpha,\beta}E_{\alpha+\beta}& \alpha+\beta\in\Delta\\
\sum_I\alpha^IH^I & \alpha+\beta=0\\
0 & \alpha+\beta \notin \Delta\end{cases}
\end{equation}
where $H^I$, $I=1\ldots 7$, denote matrix representations of the $I$-th Cartan of $E_7$, and similarly, $E_\alpha$ denote matrix representations of the generator associated to the root $\alpha\in \Delta$, with $\Delta$ the root lattice of $E_7$. Even if $\epsilon_{\alpha,\beta}=\pm 1$ can be explicitly computed, we will not need them  in our computations. Representations are constructed by systematic application of generators (associated to simple roots) to states, without anticommuting them.

A convenient way to write the roots is as vectors lying in a seven-dimensional subspace of an eight-dimensional vector space, orthogonal to $e_7+e_8$ (see for instance \cite{rootsconv}).
Namely,
\begin{eqnarray}
 (\underline {\pm1,\pm1,0,0,0,0};0,0)& & {\rm 60 \ roots} \nn \\
 \pm (0,0,0,0,0,0; 1,-1)& &  {\rm 2 \ roots} \nn \\
 \pm \frac12 (\pm1,\pm1,\pm1,\pm1,\pm1,\pm1;1,-1) &  {\rm even \ \# \ of \ - \ signs \ in \ first \ 6 \ ,} &   \rm{ 64 \  roots} \nn \ ,
\end{eqnarray}
where in the first set of roots the underline means that the two non-zero entries are at any two positions in the first six components of the vector, and in the last set of roots there should be an even number of minus signs in the first six entries. The first 60 roots together with 6 Cartan generators, generate $O(6,6)$. The next two roots, together with a Cartan
generator generate $\SLR$, while the last 64 are the adjoint $\E7$ elements in the $(\rep{32},\rep{2})$.

Positive roots are defined as
\begin{eqnarray}
 &  &   e_i\pm e_j \qquad  1\le i < j \le 6 \nn \\
 &  &  e_8-e_7 \nn \\
&& \frac12 ( e_8-e_7+\sum_{i=1}^6 (-1)^{\nu_i} e_i)  \qquad  \sum_{i=1}{\nu_i}= {\rm  even} \nn
\end{eqnarray}

A basis of simple roots is then given by
\begin{eqnarray}
\alpha_1 & =  &  (1,-1,0,0,0,0;0,0) \label{alpha} \\
\alpha_2 & =  &  (0,1,-1,0,0,0;0,0) \nn \\
\alpha_3 & =  &  (0,0,1,-1,0,0;0,0) \nn \\
\alpha_4 & =  &  (0,0,0,1,-1,0;0,0) \nn \\
\alpha_5 & =  &  (0,0,0,0,1,-1;0,0) \nn \\
\alpha_6 & =  &  (0,0,0,0,1,1,;0,0) \nn \\
\alpha_7 & =  &  \frac12 (-1,-1,-1,-1,-1,-1;-1,1) \nn
\end{eqnarray}

The corresponding dual fundamental weights  are
\begin{eqnarray}
\omega_7 & =  &  (0,0,0,0,0,0;-1,1) \equiv \omega_1^L:  {\bf 133} \nn \\
\omega_5 & =  & \frac12  (1,1,1,1,1,-1,;-2,2) \equiv \omega_2^L : {\bf 912}  \nn \\
\omega_6 & =  &  \frac12 (1,1,1,1,1,1;-3,3) \equiv \omega_3^L: {\bf 8645} \nn  \\
\omega_4 & =  &  (1,1,1,1,0,0;-2,2)  \equiv \omega_4^L : {\bf 365750} \nn \\
\omega_3  & =  & (1,1,1,0,0,0;-\frac32,\frac32) \equiv \omega_5^L: {\bf 27664}  \nn \\
\omega_2 & =  &  (1,1,0,0,0,0;-1,1) \equiv \omega_6^L :{\bf 1539} \nn \\
\omega_1  & =  &  (1,0,0,0,0,0;-\frac12,\frac12) \equiv \omega_7^L :{\bf 56}  \nn
\end{eqnarray}

We have indicated with a  superscript $L$ the corresponding weight in the  notation used in the Lie software \cite{Lie} and we have indicated for which representation they are the highest weight.

The weights of all the elements in the representations $\rep{56}$, $\rep{133}$ and $\rep{912}$ according to their $\Tsub$ assignments are

\begin{center} \underline{\bf 56} \end{center}
\begin{eqnarray}
{\bf (12,2)} & & (\underline{\pm 1,0,0,0,0,0};\frac12,-\frac12) \nn \\
& & (\underline{\pm 1,0,0,0,0,0};-\frac12,\frac12) \nn \\
{\bf (32',1)} & & \frac12(\pm1,\pm1,\pm1,\pm1,\pm1,\pm1;0,0)\qquad \textrm{odd \# of -} \nn
\end{eqnarray}

\begin{center} \underline{\bf 133} \end{center}
\begin{eqnarray}
{\bf (32,2)} & & \frac12(\pm1,\pm1,\pm1,\pm1,\pm1,\pm1;-1,1)\qquad \textrm{even \# of - in first 6} \nn \\
& & \frac12(\pm1,\pm1,\pm1,\pm1,\pm1,\pm1;1,-1)\qquad \textrm{even \# of - in first 6} \nn \\
{\bf (1,3)} & & \pm (0,0,0,0,0,0;1,-1) \quad +\textrm{ 1 Cartan} \nn \\
{\bf (66,1)}& & (\underline{\pm 1,\pm 1,0,0,0,0};0,0)\quad +\textrm{ 6 Cartans} \nn
\end{eqnarray}

\begin{center} \underline{\bf 912} \end{center}
\begin{eqnarray}
{\bf (32',3)} & & \frac12(\pm1,\pm1,\pm1,\pm1,\pm1,\pm1;2,-2) \qquad \textrm{odd \# of - in first 6}  \nn \\
& & \frac12(\pm1,\pm1,\pm1,\pm1,\pm1,\pm1;-2,2) \qquad \textrm{odd \# of - in first 6} \nn  \\
& & \frac12(\pm1,\pm1,\pm1,\pm1,\pm1,\pm1;0,0) \qquad \textrm{odd \# of - in first 6} \nn \\
{\bf (12,2)} & & (\underline{\pm1,0,0,0,0,0};\frac12,-\frac12) \nn \\
& & (\underline{\pm1,0,0,0,0,0};-\frac12,\frac12) \nn \\
{\bf (220,2)}& & (\underline{\pm1,\pm1,\pm1,0,0,0};-\frac12,\frac12) \nn \\
& & (\underline{\pm1,\pm1,\pm1,0,0,0};\frac12,-\frac12) \nn \\
& & \textrm{5 copies of }(\underline{\pm1,0,0,0,0,0};\frac12,-\frac12) \nn \\
& & \textrm{5 copies of }(\underline{\pm1,0,0,0,0,0};-\frac12,\frac12) \nn \\
{\bf (352,1)}& &  \frac12(\underline{\pm3,\pm1,\pm1,\pm1,\pm1,\pm1};0,0)\qquad \textrm{even \# of -}  \nn \\
& &\textrm{5 copies of }\frac12(\pm1,\pm1,\pm1,\pm1,\pm1,\pm1;0,0)\quad \textrm{odd \# of -} \nn
 \end{eqnarray}

\subsection{Some relevant formulas}

The action of the
adjoint representation (with parameter $A$) on the fundamental representation, decomposed in elements of $\Tsub$ is
\begin{equation} \label{adjfund}
\begin{aligned}
   \delta\lambda^{iA} &= A^i{}_j \lambda^{jA} + A^A{}_B\lambda^{iB}
      + \mukai{A^{i+}}{\Gamma^A\lambda^-} \ , \\
   \delta\lambda^- &= \tfrac{1}{4}A_{AB}\Gamma^{AB}\lambda^-
      + \epsilon_{ij}\lambda^{iA}\Gamma_A A^{j+} \ .
\end{aligned}
\end{equation}
The adjoint action on the $\rep{133}$ representation (with parameter
$A^\prime$) is given by $\delta A=[A',A]$ where
\begin{equation} \label{adjac}
\begin{aligned}
   \delta A^{ i}{}_j
      &= A^{\prime i}{}_k A^k{}_j - A^i{}_kA^{\prime k}{}_j
         + \epsilon_{jk}\left( \mukai{A^{\prime i+}}{A^{k+}}
         - \mukai{A^{i+}}{A^{\prime k+}} \right) \ , \\
 \delta  A^{ A}{}_B
      &= A^{\prime A}{}_C A^C{}_B - A^A{}_C A^{\prime C}{}_B
         + \epsilon_{ij}\mukai{A^{\prime i+}}
            {\Gamma^A{}_B A^{j+}} \ , \\
   \delta A^{ i+}
      &= A^{\prime i}{}_j A^{j+} - A^i{}_j A^{\prime j+}
         + \tfrac{1}{4}A'_{AB}\Gamma^{AB}A^{i+}
         - \tfrac{1}{4}A_{AB}\Gamma^{AB}A^{\prime i+} \ .
\end{aligned}
\end{equation}
The $\E7$ symplectic invariant is
\begin{equation} \label{sympl}
 {\cal S}(\lambda,\lambda') =  \epsilon_{ij}\lambda^{iA} \lambda'^{jB} \eta_{AB}
      + \mukai{\lambda^+}{\lambda^{'+}} \ , \\
\eeq
The projection on the $\rep{56}$ in the tensor product $\rep{912} \times \rep{133}$ is given by
\begin{equation} \label{912x133}
\begin{aligned}
\lambda^{iA}&= n_1 \, f^{jA} A^i{}_j + n_2 \, f^{iB} A^{CA} \eta_{BC} + n_3 \, \mukai{f^i{}_j{}^-}{\Gamma^A A^{j+}}+n_4 \, \mukai{f^{A+}}{A^{i+}}
+ n_5 \, f^{iABC} A_{BC} \\
\lambda^{-}&=n_6 \, \epsilon_{ij} f^{iA} \Gamma_A A^{j+}+ n_7 \, f^i{}_j{}^- A^j{}_i+n_8 \, A_{AC} \Gamma^A f^{C+}+n_9 \, \epsilon_{ij} f^{iABC} \Gamma_{ABC} A^{j+}
\end{aligned}
\end{equation}
where $n_1 ... n_9$ are some constant coefficients which we leave undetermined.

\section{Tensor structure of U-dual fluxes}
\label{912weight}

In the following tables we summarize the tensor structure of fluxes transforming in the $\mathbf{912}$ representation of $E_7$, as it can be read from their weight assignments in the $\Tsubh$ basis. We present also the corresponding weight in the $\Tsubb$ basis (c.f. Section \ref{sec:e7break}), their transformation properties under the space-time fermionic number for left-movers and the worldsheet parity operators, and the total number
of components for each flux .

States which come in 6 copies are distributed among different representations of $O(6,6)\times SL(2,\mathbb{R})$. In particular, $\bQ'^{123456}_i$, $\bQ^i
$, $\bom_i
$ and $\bom'^i
$ ($\bQ'^{123456}_i$, $\bQ^i$
, $\bP^i
$ and $\bP^{ijklm}$) transform in the $(\mathbf{12},\mathbf{2})$ in the $\Tsubh$ basis ($\Tsubb$ basis), whereas $Q'^{[i,jklm]}$, $Q^{ij}_j$, $\omega^i_{im}$ and $\omega'^i_{ij}$ ($Q'^{[i,jklm]}$, $Q^{ij}_j$, $P^{ij}_j$ and $P'^{[i,jklm]}$) transform in the $(\mathbf{220},\mathbf{2})$. Similarly, $\bP^i$, $\bP^{ijk}$ and $\bP^{ijklm}$ ($\bom_i$, $\bom'^i$ and $\bP^{ijk}$) transform in the $(\mathbf{32}',\mathbf{3})$ in the $\Tsubh$ basis ($\Tsubb$ basis), whereas $\omega^i_{ji}$, $\omega'^i_{ji}$, $L^{ijk}$ and $P^{ijkl}_i$ ($P_i^{ji}$, $P'^{[i,jklm]}$, $L^{ijk}$ and $P^{ijkl}_i$) transform in the $(\mathbf{352},\mathbf{1})$.


\begin{table}[!ht]
\begin{center}
{\footnotesize \begin{tabular}{|c|c|c|c|c|c|}
\hline
flux & $\Tsubh$ & $\Tsubb$ & $(-1)^{F_L}$ & $\Omega_P$ & Total \#\\
\hline
$\bQ'^{123456}_{1},  Q'^{[2,3456]}$& $6\times(1,0,0,0,0,0\ ;+,-)$ & $6\times(-1,0,0,0,0,0\ ;+,-)$ & $+$ & $+$ & 6+30\\
\hline
$\bQ^{1},  Q^{1m}_m$& $6\times (-1,0,0,0,0,0\ ;-,+)$ & $6\times (1,0,0,0,0,0\ ;-,+)$ & $+$ & $+$ & 6+30 \\
\hline
$\bom_{1},  \omega_{1m}^m$& $6\times (1,0,0,0,0,0\ ;-,+)$ & $6\times (-,+,+,+,+,+\ ;0,0)$ & $+$ & $-$& 6+30\\
\hline
$\bom'^
{1},  \omega'^{1m}_{m}$& $6\times (-1,0,0,0,0,0\ ;+,-)$ & $6\times (+,-,-,-,-,-\ ;0,0)$ & $+$ & $-$& 6+30 \\
\hline
$F_{12345}$& $(+,+,+,+,+,-\ ;-1,1)$ & $(+,+,+,+,+,\frac32\ ;0,0)$ & $-$ & $+$& 6\\
\hline
$F_{123}$& $(+,+,+,-,-,-\ ;-1,1)$ & $(0,0,0,1,1,1\ ;-,+)$ & $-$ & $-$& 20\\
\hline
$F_{1}$& $(+,-,-,-,-,-\ ;-1,1)$ & $(-,+,+,+,+,+\ ;-1,1)$ & $-$ & $+$& 6 \\
\hline
$\bP^
{6}, P_{m}^{6m}$& $6\times (+,+,+,+,+,-\ ;0,0)$ & $6\times (1,0,0,0,0,0\ ;+,-)$ & $-$ & $-$& 6+30\\
\hline
$\bP^
{456}, P'^{4,56}, P_{m}^{m456}$& $6 \times (+,+,+,-,-,-\ ;0,0)$ & $6 \times (-,-,-,+,+,+\ ;0,0)$ & $-$ & $+$& 20+40+60\\
\hline
$\bP^{23456}, P'^{2,3456}$& $6\times (+,-,-,-,-,-\ ;0,0)$ & $6\times (-1,0,0,0,0,0\ ;-,+)$ & $-$ & $-$ & 6+30\\
\hline
$F'^{6,123456}$& $(+,+,+,+,+,-\ ;1,-1)$ & $(-,-,-,-,-,+\ ;1,-1)$ & $-$ & $+$& 6\\
\hline
$F'^{456,123456}$
& $(+,+,+,-,-,-\ ;1,-1)$ & $(-1,-1,-1,0,0,0;+,-)$ & $-$ & $-$& 20 \\
\hline
$F'^{23456,123456}$& $(+,-,-,-,-,-\ ;1,-1)$ & $(-\frac32,-,-,-,-,-\ ;0,0)$ & $-$ & $+$ & 6\\
\hline
$H_{123}$& $(1,1,1,0,0,0\ ;-,+)$ & $(0,0,0,1,1,1\ ;+,-)$ & $+$ & $+$ &20 \\
\hline
$\omega^2_{13}$& $(1,-1,1,0,0,0\ ;-,+)$ & $(-,\frac32,-,+,+,+\ ;0,0)$ & $+$ & $-$ &60\\
\hline
$Q^{12}_3$& $(-1,-1,1,0,0,0\ ;-,+)$ & $(1,1,-1,0,0,0\ ;-,+)$ & $+$ & $+$ &60 \\
\hline
$R^{123}$& $(-1,-1,-1,0,0,0\ ;-,+)$ & $(+,+,+,-,-,-\ ;-1,1)$ & $+$ & $-$ &20 \\
\hline
$\tilde H_{123}$& $(1,1,1,0,0,0\ ;+,-)$ & $(-,-,-,+,+,+\ ;1,-1)$ & $+$ & $-$  &20 \\
\hline
$Q'^{3,3456}$& $(1,1,-1,0,0,0\ ;+,-)$ & $(-1,-1,1,0,0,0\ ;+,-)$ & $+$ & $+$  &60\\
\hline
$\omega'^{12356,56}$& $(0,0,0,1,-1,-1\ ;+,-)$ & $(-,-,-,-\frac32,+,+\ ;0,0)$ & $+$ & $-$  &60 \\
\hline
$H'^{123,123456}$& $(-1,-1,-1,0,0,0\ ;+,-)$ & $(0,0,0,-1,-1,-1\ ;-,+)$ & $+$ & $+$  &20 \\
\hline
$\hat F_{1}$& $(\frac32,+,+,+,+,+\ ;0,0)$ & $(-,+,+,+,+,+\ ;1,-1)$ & $-$ & $+$ &6 \\
\hline
$P^{23}_1$& $(\frac32,-,-,+,+,+\ ;0,0)$ & $(-1,1,1,0,0,0\ ;+,-)$ & $-$ & $-$  &60 \\
\hline
$P'^{1,12}$& $(-\frac32,-,+,+,+,+\ ;0,0)$ & $(\frac32,+,-,-,-,-\ ;0,0)$ & $-$ & $+$ &30\\
\hline
$P'^{1,1234}$& $(-\frac32,-,-,-,+,+\ ;0,0)$ & $(1,0,0,0,-1,-1\ ;-,+)$ & $-$ & $-$ &60 \\
\hline
$P_{1}^{2345}$& $(\frac32,-,-,-,-,+\ ;0,0)$ & $(-\frac32,+,+,+,+,-\ ;0,0)$ & $-$ & $+$  &30 \\
\hline
$\hat{F}'^{1,123456}$& $(-\frac32,-,-,-,-,-\ ;0,0)$ & $(+,-,-,-,-,-\ ;-1,1)$ & $-$ & $+$ &6 \\
\hline
\end{tabular}}
\caption{\small
Field strengths transforming in the $\mathbf{912}$ of $E_7$. We use the shorthand notation $\pm\equiv\pm\frac12$.} \label{tabla352}
\end{center}
\end{table}

{\small

\end{document}